# Management of quality requirements in agile and rapid software development: a systematic mapping study

Woubshet Behutiye[a*], Pertti Karhapää[a], Lidia Lopez[b], Xavier Burgués[b], Silverio Martínez-Fernández[c], Anna Maria Vollmer[c], Pilar Rodríguez[a], Xavier Franch[b], Markku Oivo[a]

[a]*M3S, Faculty of Information Technology and Electrical Engineering, University of Oulu, Oulu, Finland*

[b] *Universitat Politècnica de Catalunya, Barcelona, Spain*

[c] *Fraunhofer Institute for Experimental Software Engineering IESE, Kaiserslautern, Germany*

**Abstract**

*Context:* Quality requirements (QRs) describe the desired quality of software, and they play an important role in the success of software projects. In agile software development (ASD), QRs are often ill-defined and not well addressed due to the focus on quickly delivering functionality. Rapid software development (RSD) approaches (e.g., continuous delivery and continuous deployment), which shorten delivery times, are more prone to neglect QRs. Despite the significance of QRs in both ASD and RSD, there is limited synthesized knowledge on their management in those approaches.

*Objective:* This study aims to synthesize state-of-the-art knowledge about QR management in ASD and RSD, focusing on three aspects: bibliometric, strategies, and challenges.

*Research method:* Using a systematic mapping study with a snowballing search strategy, we identified and structured the literature on QR management in ASD and RSD.

*Results:* We found 156 primary studies: 106 are empirical studies, 16 are experience reports, and 34 are theoretical studies. Security and performance were the most commonly reported QR types. We identified various QR management strategies: 74 practices, 43 methods, 13 models, 12 frameworks, 11 advices, 10 tools, and 7 guidelines. Additionally, we identified 18 categories and 4 non-recurring challenges of managing QRs. The limited ability of ASD to handle QRs, time constraints due to short iteration cycles, limitations regarding the testing of QRs and neglect of QRs were the top categories of challenges.

*Conclusion:* Management of QRs is significant in ASD and is becoming important in RSD. This study identified research gaps, such as the need for more tools and guidelines, lightweight QR management strategies that fit short iteration cycles, investigations of the link between QRs challenges and technical debt, and extension of empirical validation of existing strategies to a wider context. It also synthesizes QR management strategies and challenges, which may be useful for practitioners.

## 1. Introduction

In the current era, which is characterized by continuous technological breakthroughs, high quality has become imperative for the acceptance of software systems. Non-secure business applications, unreliable Internet of Things platforms, or inefficient 5G-based systems can no longer survive in the context of pervasive IT. Evidence supports these claims; market prospects indicate that up to 26% of firms' IT budgets are dedicated to software quality assurance and testing, and they predict an increase to 33% in the next three years [1]. In addition, some studies have shown the impact of overlooking quality-related aspects on the costs of software development and maintenance [2–4]. For instance, Ramesh et al. [3] reported that overlooking quality-related aspects in the early stages of development resulted in degradation of system quality and required redevelopment of the whole system.

However, software quality is elusive [5]; it is not easy to define or measure. In this context, quality requirements (QRs) play a crucial role in dealing with quality. QRs are the desired qualities of a system to be developed, such as maintainability, reliability, availability, usability, and integrity [6]. Although QRs have some similarities to their functional counterpart—namely, functional requirements—they are unique in other respects, including their meaning, how they are expressed, and how they are measured. While these challenges exist regardless of the approach used to develop software, they are particularly prominent in agile software development (ASD) which entail incremental and iterative software development methods guided by agile manifesto [7], and rapid software development (RSD)[1]. ASD methods (e.g. Scrum, and XP), and RSD approaches (e.g. continuous deployment, continuous delivery, and DevOps, which extend ASD' capability by shortening the time to delivery of software), are widely adopted throughout the industry because they place focus on continuous delivery of valuable software and customer satisfaction [8,9]. In such approaches, functional requirements tend to be favored over QRs [8,10,11], leading QRs to be improperly documented [12]. Consequently, quality aspects, such as system security, performance, and usability, are often compromised [3].

Neglecting QRs introduces bottlenecks in ASD [13] and RSD processes. Specifically, Bellomo et al. [14] reveal how neglecting QRs in early phases of development may compromise performance and stability. Additionally, inadequate or missing knowledge of QRs and functional requirements often incur technical debt in ASD, leading to the need for rework, and increased maintenance costs [4]. Technical debt due to QRs results in quality issues that become difficult to test in later phases [15]. In this regard, systematic management of QRs is important.

Despite the need to better understand the management of QRs in ASD and RSD, the body of knowledge on this topic is not well structured. While there are secondary studies [10,11,16–22] that examine state-of-the-art of requirements engineering (RE) in the context of ASD, only Alsaqaf et al. [10] and Villamizar et al. [22] specifically focused on QRs. However, Alsaqaf et al. [10] did so only in the context of large-scale ASD, whereas Villamizar et al. [22] examined only how security is handled in ASD. Existing secondary studies of RSD [8,23–25] also do not specifically focus on QRs. This makes it difficult for researchers and practitioners to obtain a clear understanding of QRs and their management in ASD and RSD. Uncovering important QRs in such contexts, related challenges and gaps would be beneficial. In this paper, we aim to fill this research gap by systematically identifying, structuring, analyzing, and assessing the quality of the scientific literature on management of QRs in ASD and RSD. In short, our systematic mapping study (SMS) has the following aims:

1. To identify and classify the existing scientific research on management of QRs in ASD and RSD.
2. To assess the quality of existing empirical work on the topic in terms of research rigor and industrial relevance.
3. To identify and classify QR management strategies in ASD and RSD.
4. To identify and classify the challenges of managing QRs in ASD and RSD.

Our work contributes to the existing body of knowledge on RE in ASD and RSD. For academia, we identify important research gaps that need further investigation, such as tools and guidelines for managing QRs, lightweight QR management strategies that can be used with short iteration cycles, the link between QRs and technical debt, and empirical validation of existing strategies in a wider context. Practitioners can utilize the findings of this study to improve their understanding of QR management in ASD and RSD. Consequently, they may address and eventually overcome the quality-related issues that arise in ASD and RSD. In such scenarios, it is important they assess the relevance and applicability of the management strategies in their development contexts. The rest of the paper is organized as follows. Section 2 provides a background and related work on the topic. Section 3 presents the research methodology followed to conduct the SMS. Section 4 presents the findings of the research. Section 5 discusses the implications of the findings. Finally, Section 6 summarizes and concludes the study.

[1] RSD refers to rapid and continuous software engineering approaches that are an evolutionary step away from ASD and focus on an organization's capability to develop, release, and learn from software in rapid parallel cycles, such as those lasting for hours, days, or a few weeks [8,23,47].

## 2. Background and related work

This section presents some background information about QRs and their management in ASD and RSD. Moreover, it presents secondary studies on agile requirement engineering and RSD.

### 2.1 Quality requirements

A QR is defined as "a requirement that pertains to a quality concern that is not covered by functional requirements" [26]. QRs play a crucial role in the success of software systems. Failure to properly deal with them may result in increased cost or longer time-to-market [27,28]. The definition above, which relies upon functional requirements, has led to the widespread use of the term non-functional requirement (NFR). Mylopoulos et al. [29] defined NFRs as "… global requirements on [the system] development or operational cost, performance, reliability, maintainability, portability, robustness, and the like." Other terms, such as "extra-functional requirement" [30] or those ending in "-ilities" [31] have also been proposed in the literature, although they are not as commonly adopted as QR and NFR. The lack of consolidation of the terminology and ontological meaning of this class of requirements is well known and reported in several works [32]. In our study, we consider all the terminology to be equivalent and use QR as a representative term. As our goal is to provide a broad view of QRs, this will help ensure that we do not leave out any work that contributes to the topic under exploration.

It is important to understand which aspects are addressed by QRs in practice. In the 1970s and 1980s, researchers such as Boehm [33], McCall [34], and Roman [35] as well as organizations such as Rome Air Development Center [36] proposed their own taxonomies of QRs. For example, ISO/IEC published the 9126-1 standard [37], which was later updated and referred to as the 25010 standard [38]. The latter can be considered the most widespread method of defining, categorizing, and managing QRs. ISO/IEC 25010 comprises eight quality categories—functional suitability, performance efficiency, compatibility, usability, reliability, security, maintainability, and portability—which are subdivided into subcategories. For instance, the subcategories of functional suitability are functional completeness, functional correctness, and functional appropriateness. However, the standard does not capture some aspects well, such as those related to non-technical quality aspects (e.g., those relating to licensing or product support) [39], transversal aspects (e.g., dependability), or other characteristics that have become mainstream in modern IT systems (e.g., transparency and sustainability).

One focus of this SMS is uncovering the challenges associated with QR management. Given the elusive nature of software quality [5], several studies have addressed this topic. Kitchenham and Pfleeger [5] asked respondents to rank quality issues by selecting the three most important issues from a list. Their results show that the ranking, from most to least important, is as follows: specifying QRs objectively, setting up quality management systems, achieving operational qualities, measuring quality achievements, and agreeing with customers about what quality means. Svensson et al.'s systematic review [40] reported difficulties in eliciting QRs, dependency on the context and market when quantifying QRs, and possible misalignments among stakeholders while considering QRs. Regarding misalignments, which are closely related to the lack of ontological agreement mentioned above, Ameller et al. [41] investigated software architects' understanding of QR ontology. They found that misalignments were due to an inability to interpret some terms (e.g., "accuracy"), wrong use of terms (e.g., "friendly" instead of "usable"), and wrong definition of well-established terms (e.g., defining "maintainability" as "when something is working, we cannot make changes"). In this study and a subsequent paper, Ameller et al. [42] achieved insights into decision-making related to QR elicitation, documentation, and validation from the perspective of software architects. Our study extends the knowledge base by investigating the challenges associated with QRs in ASD and RSD, which have become more prominent in recent years. The next subsection discusses this topic.

### 2.2 Agile and rapid software development

ASD refers to software development methods that are driven by 12 principles and four values proposed in the agile manifesto [7]. It focuses on "**individuals and interactions** over processes and tools, **working software** over comprehensive documentation, **customer collaboration** over contract negotiations and **responding to change** over following a plan" [7]. Some examples of ASD methods include Scrum, XP, and feature-driven development (FDD). ASD' ability to create business value, embrace changes in requirements, and enable fast delivery of software have led to their popularity [43–45], and surveys reveal that it is now well-established in the software industry [1,9,46]. During the last few years, rapid and continuous software engineering (SE) approaches have also become popular. RSD can be understood as a step away from ASD as it aims to minimize the gap between development and deployment. It refers to an organization's capability to develop, release, and learn from software in rapid parallel cycles, which are often as short as hours or days [8,23,47]. RSD builds on well-known agile practices such as continuous integration and fast value delivery, adding new software development approaches such as DevOps, further automation, and continuous and rapid experimentation. As the software industry's interest in RSD increases, so does the number of scientific studies in this area [47].

Management of requirements in ASD and RSD is quite different compared to traditional RE approaches. For example, RE is a continuous activity that occurs over the lifetime of a system and enables requirements to change, even late in the development process. Moreover, ASD and RSD are more dynamic and involve small iterations and frequent releases. This new flexible and dynamic way of managing requirements has posed new challenges and led to many research contributions in the area, particularly concerning the management of QRs.

Agile RE has been widely explored in recent years, and interest in performing secondary studies on the topic has increased [10,11,16–18]. However, despite the significance of managing QRs in ASD and RSD [10,11,16–18], the body of knowledge on the management of QRs in ASD and RSD is not yet comprehensively understood.

### 2.3 Related secondary studies

Several secondary studies have analyzed the literature on RE in ASD. Table 1 presents these studies in chronological order. However, most of these secondary studies (except the studies performed by Alsaqaf et al. [10] and Villamizar et al. [22]) have primarily reported on functional requirements, and QRs are only a marginal topic in their primary studies (see the third column in Table 1. In addition, the search strategies of existing secondary studies are not targeted at studies on QRs. As further discussed in Section 3, one of the challenges of searching for studies on QRs is that many do not use general terms, such as QR, NFR, or RE, instead referring directly to the QR in question (e.g. security, usability). Thus, designing an accurate search string to automatically search for these studies is very laborious, as it requires many specific terms for QRs [10].

Except Alsaqaf et al. [10] and Villamizar et al. [22], existing secondary studies only superficially discuss QRs, or do not discuss them at all. Some highlight the limitations of agile methods when managing QRs and the risks of neglecting QRs [11,16–18,21], and some identify strategies for managing QRs in ASD, such as the AFFINE framework, NORMAP, and SENoR [11]. Others explicitly include QRs in the user story definition [20] and techniques for managing usability in agile [19,48]. However, no secondary studies analyze these strategies comprehensively. Villamizar et al. [22], investigated how security is handled in ASD and identified approaches that introduce artifacts and guidelines to handle security issues. Alsaqaf et al. [10], focused on identifying agile practices for engineering QRs in large-scale distributed agile projects, the challenges associated with QRs, and potential solutions to cope with the identified QR challenges in agile projects. They found 13 agile RE practices for managing QRs (e.g., face-face communication, frequent requirement prioritization, user stories, product grooming, test-driven development, and pair programming) and 12 associated challenges (e.g., lack of a requirement traceability mechanism, focus on delivering functionality at the cost of architecture flexibility, and product owner's lack of knowledge about QRs). They also found 13 emerging solutions (e.g., SENoR, ACRUM, and SCRUM frameworks). Despite the fact that Alsaqaf et al. [10] provided interesting insights, their focus on large-scale distributed agile projects and limited number of primary studies (the study is based on 60 primary studies, while our study found 156 primary studies on the topic) limit their results. Regarding the way in which QRs are managed in RSD, it has not been examined in any of the previous secondary studies of RSD [8,23–25].

Compared to other related secondary studies, our study provides more concrete contributions to the literature, including the following:

- A more complete list of primary studies (published until July 2019), on the topic (regarding not only ASD but also RSD contexts).
- A comprehensible and cohesive analysis of the literature on strategies and challenges for managing QRs in ASD. Our SMS provides a more exhaustive analysis of the literature, in which
    - QR management strategies (74 practices, 43 methods, 13 models, 12 frameworks, 11 pieces of advices, 10 tools, and 7 guidelines) are synthesized and
    - 18 categories and four non-recurring challenges are discovered.
- An appendix that lists each strategy and its original source, which can be used as an index of the scientific literature on managing QRs in ASD and RSD. This is particularly useful for practitioners.
- A mapping of QR management strategies and challenges that shows which challenges were the targets of each of the strategies.

**Table 1 Comparison between this study and related secondary studies**

| Secondary study | Goal | Number of primary studies | Search period | Research questions | Contributions to management of QRs |
|---|---|---|---|---|---|
| Our SMS | To identify and structure the state of the art of QRs management in ASD and RSD | Total:128 on QRs | Until July 2019 | RQ1. How is the research on QRs management characterized in the context of ASD and RSD?<br>RQ2. What strategies for managing QRs in ASD and RSD does the scientific literature report?<br>RQ3. What are the existing challenges of managing QRs in ASD and RSD? | Focus on ASD and RSD,<br>62 practices, 38 methods, 12 models, 9 tools, 9 advices, 7 frameworks, and 4 guidelines for managing QRs.<br>19 challenges of managing QRs. |
| Villamizar et al.[22] | To understand approaches proposed to handle security requirements in ASD | Total:21 on QRs (security) | 2005 - 2017 | RQ1: In the context of which agile methods have the approaches been proposed?<br>RQ2: Which RE phases do the approaches address?<br>RQ3: How are SR handled in each approach?<br>RQ4: What is the research type facets of the approaches?<br>RQ5: Which kind of empirical evaluation have been performed?<br>RQ6: What are the limitations faced by the identified approaches? | Focus on handling security in ASD. Identified approaches introducing new artifacts, modify agile methods, propose guidelines, tools and frameworks to handle security in ASD. Identified lack of research on SR verification and validation, limited tool support and lack of empirical evaluations as research gap? |
| Curcio et al. [16] | To have a better understanding of RE in ASD. | Total: 104<br>On QRs:10 | 2001- March 2017 | RQ1: On which RE topics are the researches on RE in ASD concentrated_<br>RQ2: What are the gaps concerning the RE in the context of ASD?<br>RQ3: What obstacles do the agile RE is facing (environment, people and resources)? | Identified neglecting QRs as a general gap. No particular focus on QRs except high-level discussion on the topic. |
| Alsaqaf et al. [10] | To identify challenges in the engineering of QRs in large-scale distributed Agile projects, the Agile practices that have contributed to the emergence of these challenges and the proposed solutions. | Total: 60<br>On QRs: 60 | Jan 2002 - April 2016 | RQ1: What are the agile practices used to engineer QRs in ALSD settings, according to published literature?<br>RQ2: What QRs challenges have been reported in agile projects, in general?<br>RQ3: What are the existing solutions to cope with neglected QRs in agile RE in general (not only in ALSD), as per RE literature? | Focus on large-scale distributed Agile settings. The study identified 13 practices for engineering requirements in general (the authors interpret these practices as ways of coping with QRs).<br>12 challenges in agile project that harm QRs; and 13 solutions proposed to overcome the identified challenges. |
| Schön et al. [11] | Synthesize the state of the art of RE in ASD, with a focus on stakeholder and user involvement. | Total: 60<br>On QRs: 13 | 2007- 2015 | RQ1: What approaches exist, which involve stakeholders in the process of RE and are compatible with ASD?<br>RQ2: Which agile methodologies, which are capable of presenting the user perspective to stakeholders, can be found?<br>RQ3: What are the common ways for requirements management in ASD? | Some problems related to the treatment of NFRs in ASD are uncovered (neglected NFRs and lack of formal acceptance tests). Identified 3 strategies for managing QRs in ASD (AFFINE framework, NORMAP and SENoR). |
| Heck and Zaidman [20] | To investigate what quality criteria exist for assessing the correctness of written agile requirements? | Total: 16<br>On QRs: 5 | 2001- 2014 | RQ1, Which are the known quality criteria for agile requirements specification? | NFRs should be included as part of the definition of User Stories. Architecture criteria should be identified for a User Story to be considered Ready. |
| Elghariani and Kama [21] | To investigate RE practices in ASD and challenges on agile RE. | Total: 22<br>On QRs: 4 | 2000- 2015 | RQ1: What are the agile RE practices?<br>RQ2: What are Agile methodology RE challenges? | Neglecting QRs identified as a general challenge. However, no particular focus on QRs (or NFRs) but only high-level discussion on the topic. |
| Medeiros et al. [49] | To investigate the QR specifications in ASD and build an explanatory model about it. | Total: 22<br>On QRs: -- | -- | RQ1: How is the quality of software requirements specification affected in agile software development? | No focus on QRs (NFRs) |
| Magues et al. [19] | To investigate integration of ASD and user-centered design. | Total: 161 (all of them on user-centered design) | Until Oct. 2015 | RQ1: What is the current state of integration between agile processes and usability? | Focus on usability and strategies for managing usability in ASD (e.g. AGILEUX, Lean UX, and XSBD). |
| Heikkila et al. [17] | To have a better understanding of RE in ASD | Total: 28<br>On QRs: 2 | Until Sept. 2014 | RQ1: What has been researched regarding requirements engineering in an agile context?<br>RQ2: What are the reported key benefits of agile RE?<br>RQ3: What are the reported problems and corresponding solutions related to agile RE? | Ignoring QRs and implementing QR relying on tacit knowledge identified as a general challenge. However, no particular focus on QRs (or NFRs). |
| Inayat et al. [18] | To map RE practices adopted and challenges faced by agile teams to understand how traditional RE issues are resolved using agile RE | Total: 21<br>On QRs: 5 | 2002- June 2013 | RQ1: What are the adopted practices of agile RE according to published empirical studies?<br>RQ2: What are the challenges of traditional RE that are resolved by agile RE?<br>RQ3: What are the practical challenges of agile RE? | Neglecting QRs identified as a general challenge. However, no particular focus on QRs (or NFRs) but only high-level discussion on the topic. |
| Medeiros et al. [50] | To investigate how RE is used in projects that adopt Agile | Total: 22<br>On QRs: -- | Until 2013 | RQ: How the requirements engineering has been conducted in projects that adopt agile methodologies? | No focus on QRs o QRs |
| Silva et al. [48] | To investigate integration of ASD and user-centered design approaches. | Total: 58 (all of them on user-centered design) | Until 2010 | RQ1: How are usability issues addressed in Agile projects?<br>RQ2: What are common practices to address usability issues in Agile methods? | Focus on usability and techniques for user-centered design in Agile. |

## 3. Research methodology

We performed an SMS following the guidelines proposed by Kitchenham et al. [51] and Peterson et al. [52]. We also adopted the snowballing guidelines proposed by Wohlin [53] to systematically search primary studies. SMS allows one to structure the evidence regarding a domain in which there is limited evidence or a broad topic [51]. Thus, in our study, SMS enabled us to structure the broad research on QR management in ASD. It was also suitable for determining and structuring the limited evidence on QR management in the context of RSD. The following subsections describe the steps of the SMS procedure performed in our study.

### 3.1 Defining research questions

The main objective of the study is to determine and structure QR management in order to understand it with respect to strategies and challenges from the viewpoint of researchers and SE practitioners in the context of ASD and RSD.

Bearing this objective in mind, our SMS provided answers to the following research questions (RQs):

- RQ1. How is the research on QR management characterized in the context of ASD and RSD?
- RQ2. Which strategies for managing QRs in ASD and RSD have been reported in the scientific literature?
- RQ3. What are the existing challenges associated with managing QRs in ASD and RSD?

Through the first RQ, we aim to identify and structure the QR management literature, by focusing on bibliometric of the primary studies. We also aim to investigate the quality of empirical primary studies, in order to identify gaps regarding quality of primary studies. With the second RQ, we aim to identify and classify existing QR management strategies in the literature that researchers and practitioners may utilize to further investigate and address QR challenges, respectively. With the third RQ, we aim to identify open research questions and challenges related to management of QRs in order to establish a roadmap for future work.

### 3.2 Conducting the search

This section describes the search process, which involves two steps: identifying the start set of papers and snowballing (i.e., performing iterative backward and forward snowballing on the start set of papers to determine the primary studies). Fig. 1 describes the search process.

A key characteristic of this SMS is its search strategy. The most popular search strategy among secondary studies involves performing search queries in databases by using search strings, as discussed by Kitchenham et al. [51]. The idea is to form a strong search string that applies to all the research covering a specific topic and use it in several databases to obtain a holistic and unbiased view of the topic. Therefore, the challenge is to create a proper search string that allows for an unbiased secondary study. This search strategy is inconvenient for topics for which it is not possible to formulate a strong search string (e.g., many terms used for the same concept, which was a challenge in our case) or primary studies are not included in databases (e.g., grey literature, industrial results) [54].

An alternative is to use a start set of papers (i.e., an initial set of relevant papers on the topic that are included to achieve an unbiased view of the research) and apply a rigorous snowballing process to identify more studies. The snowballing search strategy involves iterative backward and forward snowballing on the start set of papers to create a list of primary studies. Backward snowballing uses the reference list of a paper to identify more studies, whereas forward snowballing uses the studies citing a given paper to identify additional studies [53]. The problems associated with this search strategy are the time required to perform a rigorous snowballing process and identifying a good start set of papers.

The field of QR management in ASD and RSD, as reported by previous secondary studies, is characterized by the following:

- There is a very strong set of studies investigating RE in ASD. Indeed, existing secondary studies have already looked for primary studies on RE in ASD using search queries in databases (see Section 2.3), and they have considered both functional requirements and QRs. Therefore, the primary studies identified by these secondary studies can be used to build a tentative start set of papers for snowballing.
- Defining an accurate search string is very challenging because primary studies do not consistently use terms such as "QR", but terminology related to the specific QR (e.g., "maintainability" and "security"). Alsaqaf et al. [10], who used the search string "*((Agile OR agility) AND (Requirements OR non-functional requirements OR non-functional OR quality requirements OR quality attributes OR quality))*," faced by this issue: "*We assume that there may well be many more publications that took such a specific QRs perspective, however we did not hit them in Scopus because our search string was not designed for this purpose*."

- Defining search strings maybe difficult while considering terms such as "agility" as synonyms of ASD. Other disciplines use the term "agility" which creates a lot of noise in the search process. Thus, an additional filter must be applied when using databases, which may lead one to miss relevant research that is not classified properly.

For these reasons, we believe that a method involving a starting set of papers based on previous secondary studies on RE in ASD and RSD as well as snowballing was more suitable and much stronger than one involving a limited search string. Indeed, the search process with snowballing found many new primary studies compared to related secondary studies.

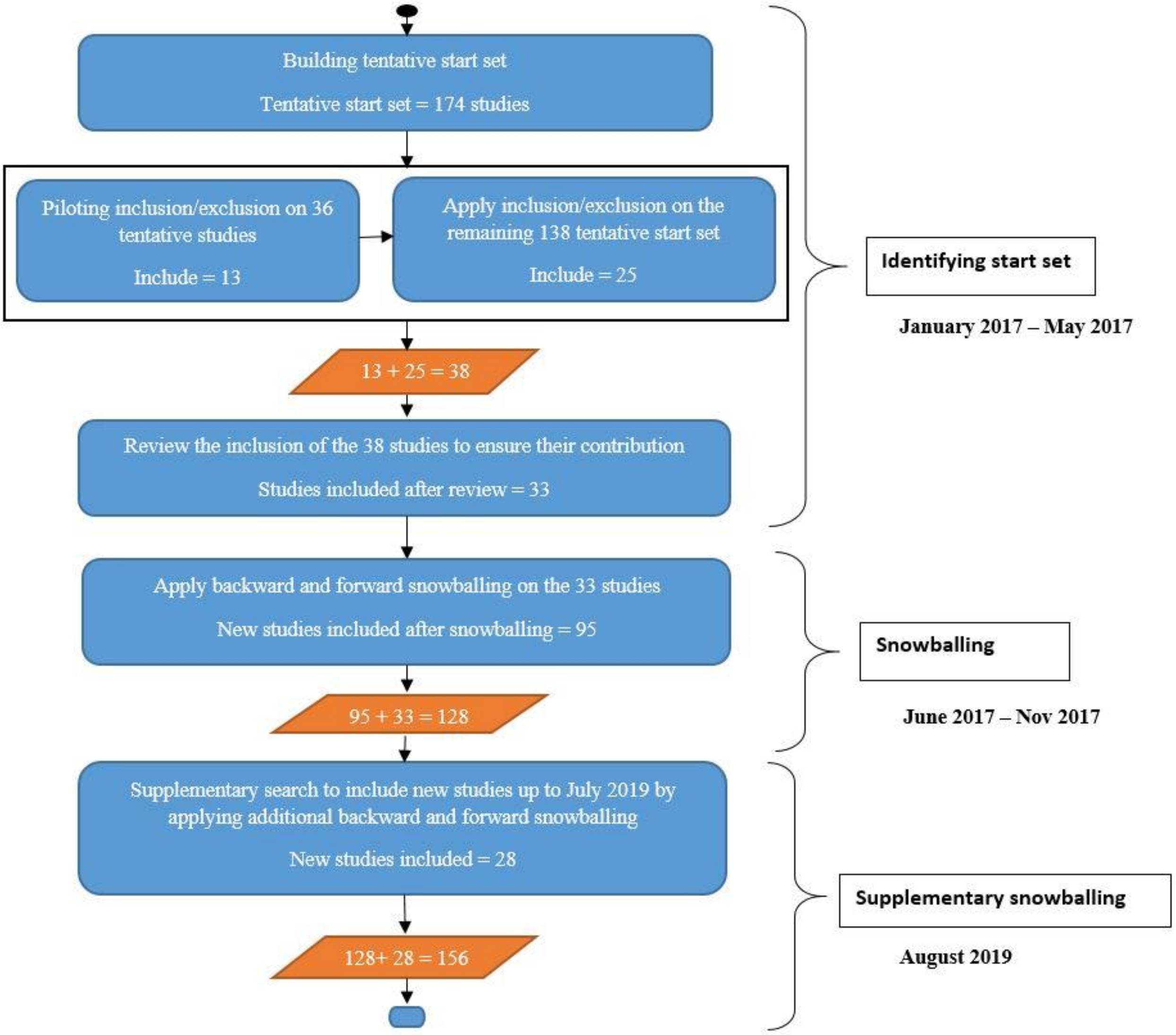

**Fig. 1. Conducting the search**

### 3.2.1 Identifying the start set

In order to identify the start set of papers, we performed the following steps: (1) build a tentative start set, (2) set and pilot inclusion/exclusion criteria on the tentative start set, (3) apply inclusion/exclusion criteria to create the start set and (4) Review included papers to ensure their contribution to the topic.

1. *Build the tentative start set*

We used the primary studies identified in secondary studies on RE in ASD [11,17,18,20] and RSD [8,23,24] to build a tentative start set. We collected 195 primary studies from the aforementioned 7 secondary studies. Of these, we excluded 21 exact duplicates. As a result, the tentative start set was comprised of 174 studies.

2. *Set and pilot inclusion/exclusion criteria on the tentative start set*

We set the inclusion and exclusion criteria to select the start set of papers from the tentative start set.

The inclusion criteria were as follows:

- Scientific studies that contribute to the body of knowledge of QRs in ASD or RSD (i.e., those that discuss different aspects of QRs in the context of ASD or RSD).
- Scientific studies published in journals, workshops, conferences, and book chapters
- Peer-reviewed publications and book chapters
- Publications written in English

Exclusion criteria:

- Scientific studies that do not contribute to the body of knowledge of QRs in ASD or RSD (e.g., those that only mention or discuss QRs in a context other than ASD or RSD or those that only mention specific practices of ASD or RSD without further discussion of QRs)
- Books, blogs, short papers, tutorials, and so on
- Secondary studies (i.e., SLRs and SMSs)
- Exact duplicates of another study
- Publication in a language other than English

Six researchers piloted the inclusion/exclusion criteria in two rounds with 36 randomly selected studies from the tentative start set. In the first pilot round, all six researchers agreed to include 2 papers and exclude 8 papers. However, their decisions for the other 26 studies differed. Specifically, there were differences in the way the inclusion criteria were understood (e.g., which papers contribute to the body of knowledge of QR management in ASD or RSD and how short papers were defined). We discussed and resolved the issues and then updated the inclusion criteria to clarify the misunderstandings. Then, we piloted the inclusion and exclusion criteria again. In the second pilot, the researchers agreed to include 13 studies and exclude 23 studies.

3. *Apply the inclusion/exclusion criteria to create the start set*

After validating the inclusion/exclusion criteria, we divided the remaining 138 studies into 3 groups of 46 studies. In each group, two researchers applied the inclusion/exclusion criteria to the 46 studies individually and compared their results. This helped minimize researcher bias. When the two researchers disagreed regarding the inclusion of a paper, we conducted a voting workshop [55], where all six researchers reviewed the paper, discussed and voted to resolve the disagreement.

During this step, 24 studies were included from all three groups. There were four studies that provoked disagreement between the two researchers. Consequently, we had a voting workshop where all six researchers reviewed the 4 studies, and decided to include only one study. Thus, in total, we included 25 studies.

4. *Review inclusion of papers to ensure contribution*

Before moving on to the snowballing process, we wanted to make sure that the 38 included studies (13 from the piloting step and 25 from the current step) clearly contribute to the research topic. Therefore, we randomly distributed and reviewed the 38 studies by mapping their findings to our research questions. This led us to revise our decisions regarding the inclusion of five studies. Therefore, the start set of papers used for snowballing included 33 studies.

#### 3.2.2 Snowballing

In this study, we applied the snowballing guidelines for systematic reviews proposed by Wohlin [53] to search for and identify primary studies. We performed backward and forward snowballing on the start set (33 studies) to identify additional primary studies. For this purpose, we divided the 33 studies into three groups. In order to facilitate the snowballing process, we used a shared spreadsheet that allowed the researchers to track their progress. As we were working in a distributed environment, this helped ensure that no redundant work was done; researchers could avoid working on the same paper simultaneously, and they did not need to wait for their colleagues' results to continue the snowballing process.

We used Google Scholar citations to perform forward snowballing and the reference lists of studies to conduct backward snowballing. Initially, the search was performed between June 2017 and November 2017. During this period, we included 95 additional studies, 69 identified from forward snowballing and 26 identified from backward snowballing. Thus, in total, we found 128 primary studies (33 in the start set and 95 from snowballing). Later on, we performed a supplementary snowballing search (forward and backward snowballing) to include primary studies published between November 2017 and July 2019 and included 28 more studies. Therefore, finally we found a total of 156 primary studies, see here[2] for the full list due to space limitations.

## 3.3 Data extraction

We used an Excel spreadsheet to extract data. We defined the properties of data that address our research questions regarding bibliometric, QR management strategies, and challenges. Appendix A presents a detailed description of the data properties.

Before starting data extraction, we performed two rounds of piloting for our data extraction form. The first round, in which five primary studies were included, enabled us to improve our initial data extraction form and clarify our understanding of the data properties. Then, we performed a second pilot round including five new primary studies. In this round, we achieved more than 75% agreement on the extracted data properties. Finally, we performed data extraction on 156 primary studies.

For data extraction, we divided the 156 studies into three groups. For each group, two researchers individually performed data extraction on half of the studies. Additionally, each of the six researchers crosschecked the data extracted by the other researcher in his or her group with the findings of a primary study. This step helped ensure correct extraction of data and minimize human error.

## 3.4 Quality assessment

We adapted Ivarsson and Gorschek's [56] industrial quality assessment model to assess the quality of empirical primary studies (i.e., their research rigor and industrial relevance). According to Ivarsson and Gorschek [56], research rigor is evaluated by examining the study design, study context, and validity. We evaluated the reporting of these aspects using the following rating: weak (0), medium (0.5), and strong (1). For instance, while assessing validity aspect of research rigor of a primary study, if the study provided detailed description of threats to validity of the study (e.g. internal, external, construct, and conclusion) and ways to mitigate the threats we scored 1 for the validity aspect. In cases, where the validity was described to some extent (e.g. described without details of mitigation), we scored 0.5 and when there was no description of validity we scored 0. We calculated the total research rigor as the sum of the values of all aspects (study design, study context and validity). Industrial relevance refers to the realism of the study environment in terms of the subjects, the context, the scalability, and the research method reported in the empirical studies. We evaluated these aspects using the following rating: contributing to relevance (1) and not contributing to relevance (0). For instance, while assessing subject aspect of industrial relevance of a primary study that proposes QR management strategy, if the strategy has been evaluated with practitioners who are intended users of the strategy, we scored 1. However, if the management strategy proposal was evaluated with students, which were not the intended users, we scored 0. We calculated the total relevance of a study as the sum of the values of all aspects (subjects, context, scalability, and research method). Appendix A includes detailed information on how each aspect of the research rigor and industrial relevance is examined in the quality assessment. As it was necessary to extract the data properties to assess the quality of empirical studies, we performed the quality assessment in parallel with data extraction. Therefore, we piloted the quality assessment and the data extraction form together.

## 3.5 Data analysis and synthesis

We performed both frequency and qualitative analysis to classify and synthesize the extracted data. Frequency analysis was applied to determine annual publication trends and the distribution of different types of QRs. Qualitative analysis was used while determining the QR management strategies and challenges.

We adapted the taxonomy of the Industry Classification Benchmark (ICB) [57] to classify the domains reported in the primary studies. ICB provides a detailed structure to classify industries and their sectors. We chose the super-sector level in the taxonomy (the second level in the hierarchy), as the first level is too general and lower levels would cause great dispersion. We assigned values based on what was reported by the authors of the primary studies.

Initially, we classified the QR management strategies into advice, frameworks, guidelines, methods, models, practices, and tools, as reported in the primary studies. However, when the studies did not explicitly report the aforementioned categories, we adapted the research contribution classification proposed by Shaw [58] and the method classification proposed by Brinkkemper [59], and also introducing our own definitions for guidelines, practice, and advice to guide our classification of QR management strategies.

[2] https://docs.google.com/spreadsheets/d/1FbITbwh1xat11ILVrJpU9Jv1F9awGEbmtAP1AQI7f-g/edit?usp=sharing

We also used thematic analysis [60] to synthesize the themes identified among the challenges of managing QRs in ASD and RSD. Thematic analysis is based on the conceptual relationships within coding and recurring patterns. We first applied open coding to extracted challenges, then categorized related codes to form themes, and further refined the themes to identify higher-order themes (i.e., categories) of challenges. For instance, under the *time constraint due to short iteration cycles* theme, which is identified as one challenge, we have challenges such as *S20 "But in agile approaches, the time constraints do not allow to follow usability approaches, so it is often neglected and the whole focus is on the development of a running system", S31 "The main reason for usability evaluation not being conducted more frequently is time constraint", S60 "As the Scrum release cycle is too short, there is not enough time for development team to address security requirements for each release*.", that have been linked to form the theme.

### 3.6 Threats to validity

We applied various mitigations to minimize the construct, internal, external and conclusion validity threats [61] in our study and thus improve reliability of our study. Construct validity entails identifying and applying appropriate operational measures for the concepts under study. We applied an SMS protocol with all relevant details (i.e., concepts, research objective, research questions, search method, study selection, data extraction, analysis), which was reviewed by researchers to guide the SMS. This helped mitigate threats that could arise from imprecise description of the SMS setting, misunderstanding of concepts (e.g. QRs, ASD and RSD concepts) and inappropriate RQs.

Internal validity involves determining a causal relationship between factors within the context of a given study. We applied snowballing search to retrieve as many relevant primary studies as possible. In snowballing search, a potential threat comes from difficulties in identifying good start set of papers (i.e., relevant and adequate number of start set of papers) for snowballing, which has been claimed to be problematic for systematic reviews using snowballing search strategy [53]. We minimized this threat by using existing secondary studies of RE in ASD and RSD to identify the initial tentative start set (174 papers). These studies structured and synthesized knowledge on RE in ASD and RSD, and they were published in varying scientific venues and in different years, which increased their value in the start set. Another threat to internal validity comes from study selection bias. We piloted the inclusion/exclusion criteria on 36 studies with six researchers to mitigate this threat. This helped to clarify differences and build a common understanding of the inclusion/exclusion criteria. Additionally, we performed researcher triangulation in order to minimize researcher bias when selecting the primary studies.

External validity defines the applicability of the findings of a study to other contexts. We included only peer reviewed primary studies. The excluded non-peer reviewed literature may affect the generalizability of our findings.

Conclusion validity shows the extent to which the procedures of a study are repeatable with the same result. We applied an SMS protocol, reviewed by all researchers, to guide the study. We mitigated conclusion validity threat that may arise from data extraction bias by piloting the data extraction in two rounds with the six researchers who participated in the data extraction phase. We also performed a review process in which one researcher reviews the data extracted by another researcher. This helps minimize errors that can happen in the data extraction spreadsheet and ensure that the extracted data is correct.

## 4. Results and analysis

In this section, we present the results and discuss how they relate to our research questions. Section 4.1 provides an overview of the research on the topic. Section 4.2 presents the findings regarding QR management strategies, and Section 4.3 presents the findings regarding the challenges of managing QRs. Finally, Section 4.4 maps the QR management strategies with the challenges.

### 4.1 Overview of research on management of QRs in ASD and RSD (RQ1)

We identified 156 primary studies published between 2002 and 2019 and classified them in terms of ASD/RSD context, author affiliation, venue, type of research, ICB domain, and reported QR type.

*Publication trend:* About 62% of the primary studies (97) were published from 2013 onwards (see Fig. 2 (a)).The highest number of studies (22) were published in 2017.

*ASD and RSD contexts*: Among the 156 primary studies, 92% (143) were conducted in an ASD context. Six percent (13 studies [S17, S38, S53, S77, S83, S96, S98, S106, S137, S142, S144, 145 and S152]) were conducted in an RSD context.

*Authors' affiliation*: As shown in Fig. 2(a), authors in academia conducted about 74% of the primary studies (116), 15% (23) were academia-industry collaborations, and about 11% (17) were conducted entirely by the industry. Of the latter 17 studies, five were published from 2011 onwards.

*Type of research*: We investigated the type of research in terms of empirical, theoretical, and experience report distributions. About 68% of the primary studies (106) were empirical, while 22% (34) were theoretical. Additionally, 10% (16) were experience reports. Analysis of the empirical studies' research methods shows that case studies were the most popular (49 case studies and 8 multiple-case studies). Moreover, we found 12 empirical studies applying interviews, 8 applying experiments, 11 applying surveys, 4 applying observations, and 6 applying action research. Five primary studies applied mixed methods (i.e., a combination of interviews and observations with surveys and questionnaires), while another 3 applied quantitative analysis (i.e., statistics analysis) of data collected from real-life projects.

*Venue type:* We classified the primary studies' publication venues into conferences, journals, workshops, and book chapters. We found that 62% of the primary studies (97) were conference papers and that 31% (48) were journal publications. Six percent (9) were workshop papers, while 1% (2) were book chapters.

*Domain classification*: We found that 58% of the primary studies (91) reported the domains in which they were conducted. We classified the domains following the ICB [57] and found that technology and industrial goods and services were the top two domains in our study. The technology domain includes subsectors such as computer services, the Internet, and software and computer hardware, and 36 primary studies fall into this category. Industrial goods and services, the second most common domain, includes subsectors such as business support services and industrial suppliers, which embrace most information systems. There are 33primary studies in this category. Detailed information about the classification of the domains is accessible here[3].

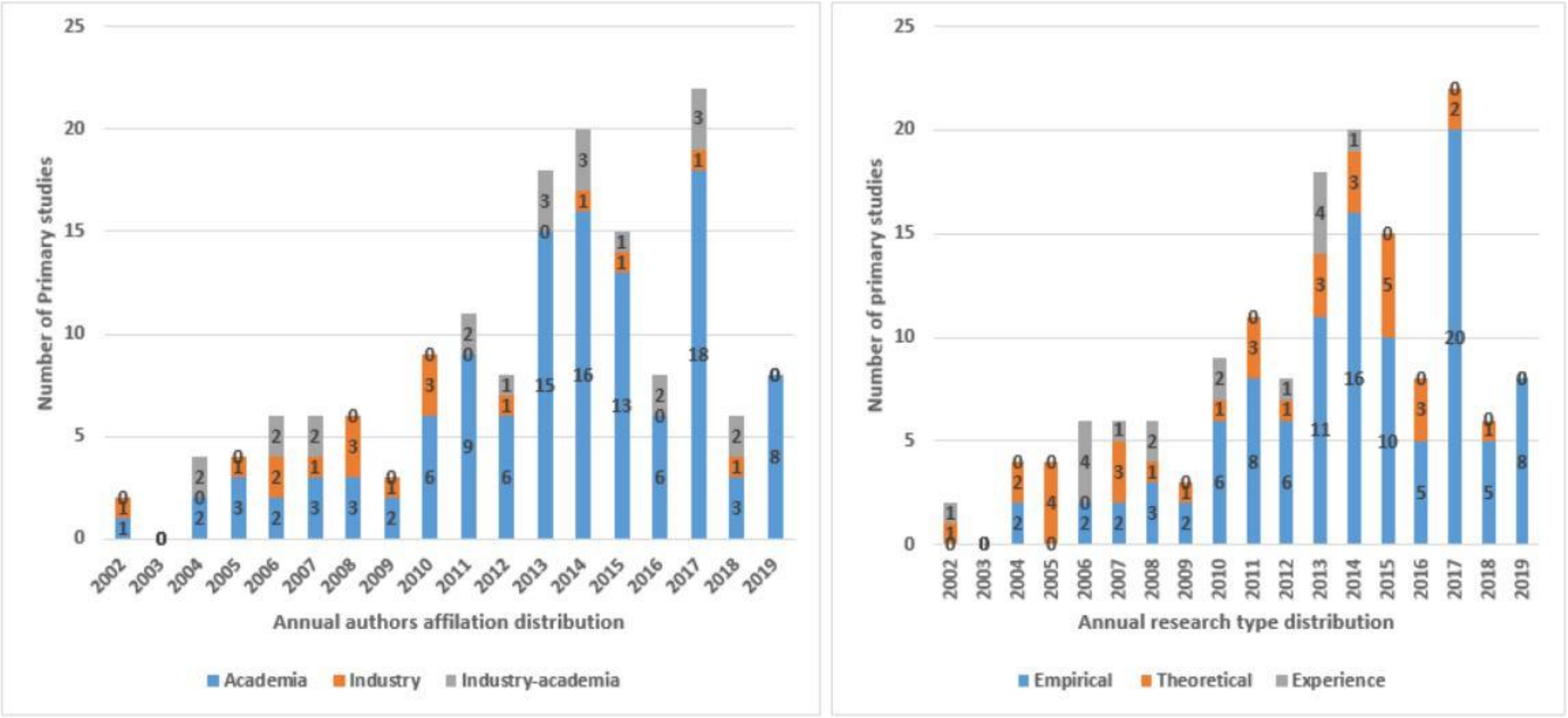

**Fig. 2(a) Annual distribution of primary studies based on authors' affiliation; (b) Annual distribution of primary studies based on research type**

*QRs types:* About 93% of the primary studies (145) reported 68 different types of QRs, while 7% (11) did not differentiate between types of QRs. We plotted the 36 QR types that were reported at least twice in Fig. 3. Security was the most popular, while performance and usability were the second and third most widely mentioned QRs, respectively. Fair attention was also paid to maintainability, reliability, scalability, and availability. We also found 32 non-recurring types of QRs. Detailed information about the distribution of QR types is accessible here[4].

[3] https://docs.google.com/spreadsheets/d/1-zNtvyIx5z9iIG4gyrzMwmyBzQp9fUe1_La9oIjmUlA/edit?usp=sharing
[4] https://docs.google.com/spreadsheets/d/1F__GPfO_6YrSbSbS-v6zvwMbAR3oflO9huWudDJFpEs/edit?usp=sharing

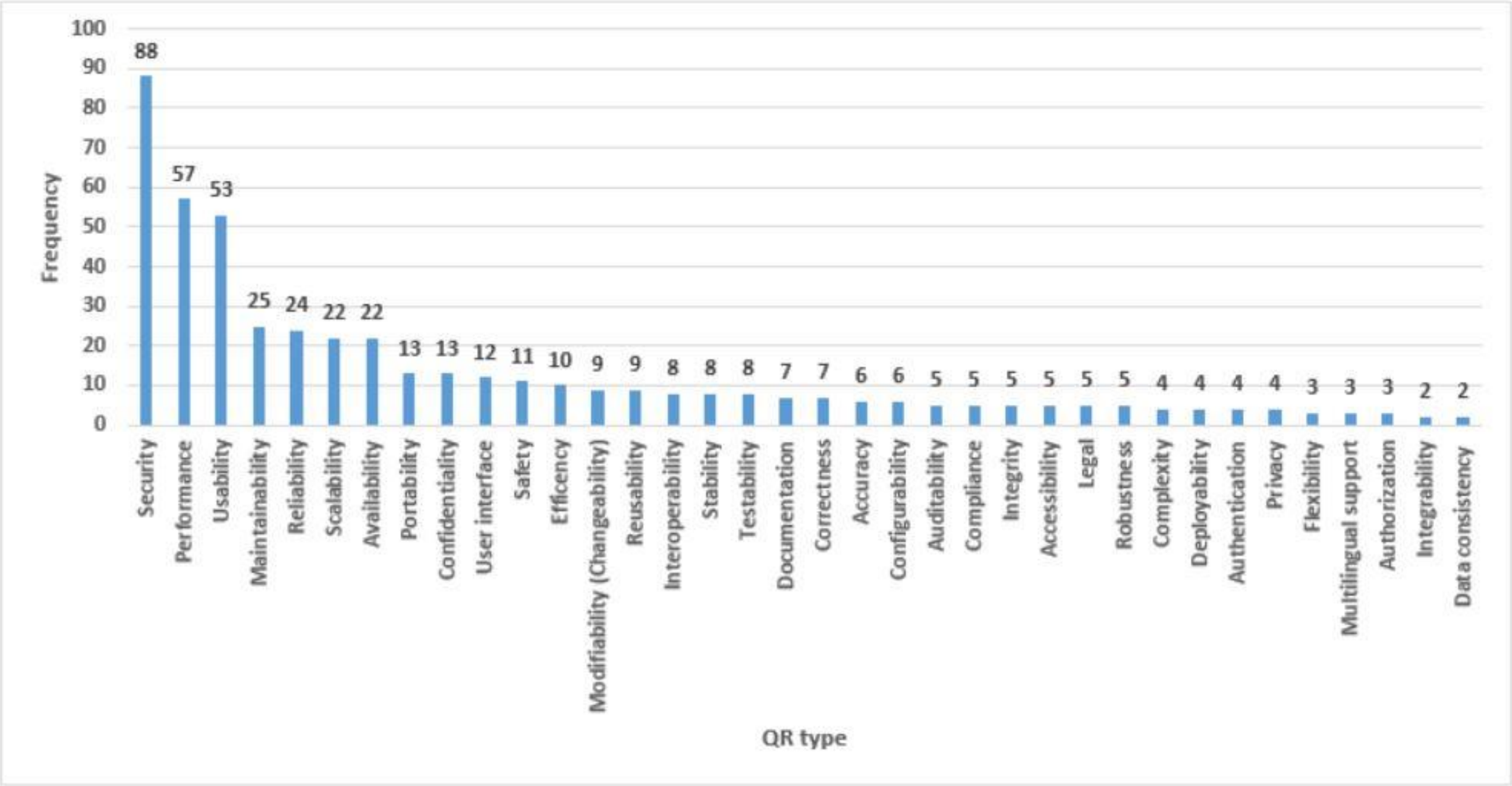

**Fig. 3. Distribution of QR types**

*Quality assessment of empirical studies:* We adapted Ivarsson and Gorschek's [56] research rigor and industrial relevance assessment model to assess the quality of 106 empirical primary studies (see Section 3.4). The bubble plot shown in Fig. 4 combines evaluations of research rigor and industrial relevance. The bubbles' size represents the number of primary studies with the respective values of research rigor and industrial relevance.

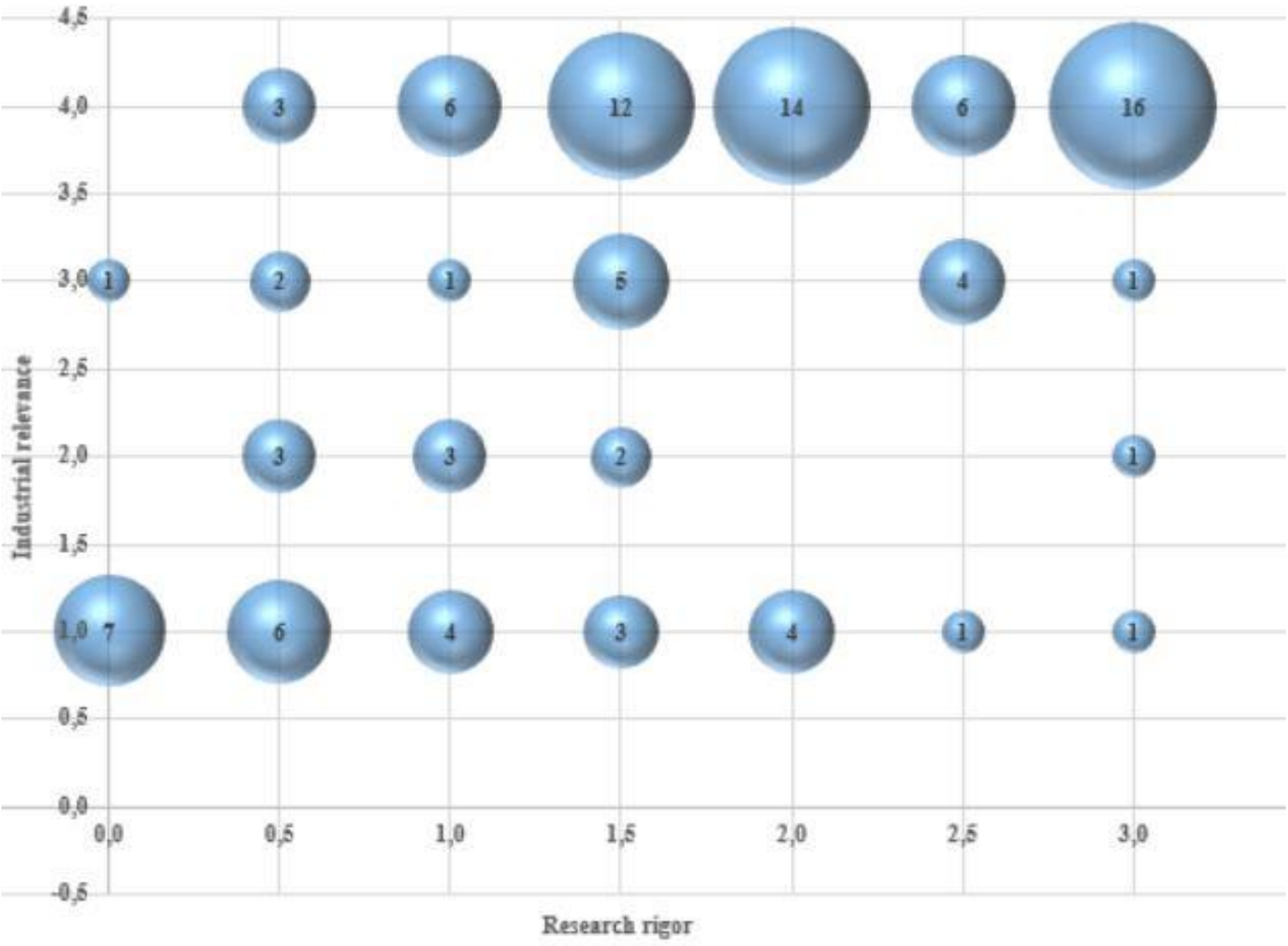

**Fig. 4. Research rigor and industrial relevance plot**

The figure shows that 61 studies lie in the upper right corner and have a research rigor value of ≥ 1.5 and an industrial relevance value of ≥ 2. Among these studies, 16 have the highest research rigor value (3) and the highest industrial relevance value (4). Studies with high research rigor and industrial relevance enable easy transfer of knowledge and offer opportunities for replication. All primary studies have an industrial relevance of ≥ 1, and 80 have industrial relevance value of ≥ 2. These results show that industrial relevance aspects were well considered in the empirical primary studies. In total, 36 primary studies have a research rigor of < 1.5. The 13 primary studies that lie in the bottom left corner have a research rigor value of < 1 and industrial relevance value of 1.

## 4.2 Strategies for managing QRs in ASD and RSD (RQ2)

In total, 143 primary studies, 93 of which were empirical, reported strategies for managing QRs in ASD and RSD. The QR management strategies helped engineer and monitor QRs (e.g., by eliciting, classifying, reasoning, prioritizing, measuring, and planning QRs) in ASD and RSD. We classified the strategies into methods, practices, models, frameworks, advices, tools, and guidelines, as shown in Table 2. Appendix B, provides a detailed list of the strategies.

**Table 2 Distribution of primary studies reporting QR management strategies (studies reporting management strategies in an RSD context are in bold font)**

| QR management strategy | Primary studies | Number of primary studies | Number of reported strategies |
|---|---|---|---|
| Practice | S7, S8, S9, S15, S17, S18, S19, S20, S31, S32, S35, S36, S37, S39, S40, S41, S56, S58, S63, S69, S78, S79, S80, S82, S95, S98, S101, S104, S106, S109, S112, S113, S119, S122, S124, S125, S128, S129, S130, S132, S142, S143, S154 | 43 | 74 |
| Method | S2, S10, S11, S23, S24, S25, S42, S43, S44, S46, S48, S49, S52, S57, S59, S64, S67, S70, S73, S74, S75, S81, S82, S85, S86, S87, S88, S89, S91, S92, S93, S94, S97, S99, S108, S110, S111, S116, S118, S120, S121, S134, S135, S149, S153, S155 | 46 | 43 |
| Model | S5, S6, S28, S30, S34, S54, S55, S60, S68, **S83**, **S96**, S100, S105, S131 | 14 | 13 |
| Framework | S4, S26, S27, S47, S51, S66, S102, S117.S136, S146, S150, S151, **S152** | 13 | 12 |
| Advice | S3, S13, S21, S30, S37, S45, S61, S84, S103, S133, **S137** | 11 | 11 |
| Tool | S5, S13, S14, S48, S50, **S53**, S55, S62, S79, S114 | 11 | 10 |
| Guidelines | S12, **S38**, S81, S127, S132, S140, S156 | 7 | 7 |

### *4.2.1 Practices*

We identified 43 primary studies that report 74 practices for managing QRs in ASD and RSD. Among these 43 studies, 21 report practices for managing QRs without specifying the type of QRs they concern. Thirteen studies [S7, S8, S9, S15, S36, S58, S78, S95, S98, S101, S106, S122, S143] report practices for managing security, six studies [S36, S112, S119, S124, S125, S154] report practices for managing usability, two studies [S32, S37] report practices for managing performance, and one study [S128] reports a practice for managing stability and robustness.

Examples of practices for managing QRs include prototyping with a focus on quality attributes [S19, S20], agile requirement engineering with prototyping [S80], quality attribute workshops to elicit QRs [S82], experimentation with A/B testing [S98], definition of ready of user stories that consider QRs [S104], definition of a dedicated architecture for QRs [S112], the incremental architecture approach [S63], and delivery stories [S39].

Two relevant focuses of many practices are introduction of specific roles and artifacts and addition of means to improve the communication of quality aspects among different stakeholders. Examples of specific roles include security expert [S7] or security master roles [S9] to address security in ASD. New artifacts such as misuse stories [S78], abuser stories [S101], and security backlogs [S9, S95] are also proposed to improve security in ASD. Similarly, we found practices that assign a usability expert role [S113] and utilize artifacts such as UserXstories [S36], discount usability [S119], and shadow backlogs [S125] to manage usability in ASD.

Regarding communication of quality aspects among different stakeholders, the studies reported practices of explicitly discussing QRs among customers and agile team leaders [S31, S41], discussion of QRs at early stages [S31], increasing awareness of QRs through education and training (e.g., on security requirements [S7, S58, S143]), and performing usability training sessions for agile designers [S125].

### *4.2.2 Methods*

We found 46 studies reporting 43 methods of managing QRs in ASD. The methods present techniques and approaches to support elicitation, reasoning, prioritization, and integration of QRs, and specific QRs (e.g. security, usability) in ASD. In this group of studies, 27 report methods for addressing QRs in general, 12 [S11, S23, S24, S46, S57, S75, S86, S92, S99, S108, S118, S149] report methods for addressing security, 6 [S81, S85, S91, S97, S116, S153] report methods for addressing usability, and 1 [S155] report method to address safety in ASD.

Examples of methods include the Non-functional requirements Elicitation, Reasoning, and Validation in Agile Processes (NERV) method [S42], the Method for Elicitation, Documentation and Validation of Software User

Requirements (MEDOV) [S43, S44], Non-functional Requirements Modeling for Agile Processes (NORMAP) [S48, S49] and Q-Rapids to elicit, assess and document QRs in ASD [S134]. For instance, Domah et al. [S42] used the NERV method, which applies different artifacts to help manage QRs in the early stages of ASD. It uses agile user story cards to capture functional requirements, a NFRusCOM card to handle QR information and capture QR elements, an NFR trigger card to elicit QRs, and NFR reasoning taxonomy to classify QRs. It also uses NFR quantification taxonomy to quantify the validation criteria of QRs and the NERV Agility Index score (NAI) to determine the degree of agility of each functional requirement and QR. The authors report that NERV effectively elicited QRs in two case studies.

Examples of methods involving specific QRs are extended agile security testing (EAST) [S46], the Security-Enhanced Agile Software Development Process (SEAP) [S10, S23], and U-Scrum [S116]. Erdogan et al. [S46] used EAST to address security issues in web applications developed with ASD. The method combines agile security testing practices with Scrum. It effectively reduced the amount of time spent on testing security and was 95% more effective at identifying vulnerability issues compared to ad hoc security testing methodologies used in ASD. Singh et al. [S116] used U-Scrum, a method that introduces the role of usability product owner, who focuses on usability and leads the user experience vision for the product, to Scrum. The method is based on interaction between the main product owner and the usability product owner, and it applies a project plan, user experience vision, and backlog structure to examine usability. In their experience report, the authors claimed that applying U-Scrum improved the usability of developed products by focusing on usability concerns during development [S116].

#### *4.2.3 Models*

We found 14 primary studies that report 13 models supporting activities such as identification, prioritization, and specification of QRs in ASD and RSD. Of these studies, eight report models for managing QRs in general, two [S54, S60] report models that focus on security, two [S28, S29] report models focusing on usability, one [S68] reports a model focusing on performance, and one reports a model focusing on reliability [S83]. The studies use agile NFRS traceability process model [S5, and S54], continuous integration visualization technique (CIVIT) [S96], and Story Card Maturity Model (SMM) [S100] to aid identification of QRs, estimation and release planning activities in ASD, and one uses US-Scrum, a hybrid model based on Scrum and FDD aimed at addressing security, usability and correctness [S105]. The software performance requirements evolution model (PREM) was used to specify and validate performance requirements in ASD [S68], and the enhanced Scrum model [S60] and conceptual secure feature-driven development (SFDD) method [S54] were also included in this category.

Only five primary studies empirically evaluated the models [S34, S54, S60, S83, S96]. For instance, Nilsson et al. [S96] reported that the CIVIT model helped visualize which QRs were tested, the degree to which they were tested, and when they were tested. According to the authors, the companies in the study were able to visualize the end-to-end process of testing activities related to QRs and hence improve QR testing activities and discussions [S96].

#### *4.2.4 Frameworks*

We found 13 primary studies reporting 12 frameworks for managing QRs in ASD and RSD. These frameworks present conceptual structures of practices and artifacts for managing QRs. Six studies [S26, S27, S47, S51, S150, S152] address QRs in general, four [S66, S102, S117, S136, ] focus on security, and three [S4, S146, S151] focus on usability. The frameworks included a user-centered design (UCD) framework to address usability in Scrum [S4], RE-KOMBINE to support lightweight requirements engineering process in ASD [S47], NORVIEW to visualize and plan QRs [S51], and Secure Scrum framework that extends Scrum with S-tags and the Agile Framework For Integrating Nonfunctional Requirements Engineering (AFFINE) [S26, S27]. Pohl et al. [S102] applied Secure Scrum to introduce security tagging in backlogs and identify security activities and appropriate security testing methods to help developers manage security in ASD. Similarly, AFFINE provided a reference architecture to support integration of QRs in ASD in two studies [S26, S27]. According to Bourimi et al. [S26], AFFINE helped consider QRs early and sufficiently in ASD. Shahin et al. [S152] proposed a conceptual framework to assist architecting for continuous delivery, by considering operational and QR aspects.

#### *4.2.5 Advice*

Eleven primary studies reported nine pieces of advice for managing QRs in ASD and RSD. Among these, six primary studies [S3, S13, S21, S30, S103, S137] provided advice for managing security, two [S37, S45] provided advice for usability, one [S84] provided advice for performance, one [S61] for managing QRs in general and one [S133] provided advice for safety. We found various suggestions for managing security in ASD, such as improving team members' awareness of security [S3, S37], including a chief security officer [S37], and performing systematic threat analysis and risk assessment to identify which security aspects to focus on in ASD [S13]. Cajander et al. [S30] suggest achieving a clear perspective on usability by project management and Scrum team and developing the usability skills of team members to help manage usability in Scrum. We also found studies that suggested applying iterative security assurance methods in ASD [S21], introducing usability evaluations to a sprint review [S45], clarifying QRs early in development [S61] and including safety team member in ASD to address safety [S133].

#### *4.2.6 Tools*

Eleven primary studies reported ten tools for managing QRs in ASD and RSD. The tools varied in terms of the type of QR they address and the activities that they aim to support. Four studies reported tools focusing on security [S5, S13, S14 and S55], three [S48, S50, S147] reported tools for managing QRs in general, two [S62, S79] reported tools focusing on usability, and two others [S53, S114] reported tools focusing on performance. For instance, the Sagile tool was used to support traceability and management of QRs in ASD [S5], and the NORMATIC tool was used to aid modeling of QRs in the requirement-gathering and analysis phase [S48, S50]. We also found tools such as POLVO, which was used to address usability in ASD by enabling prototyping, and testing usability [S62]; Perflab, which was used to monitor performance issues in continuous deployment [S53]; SQUISH, which was used for automating usability testing in ASD [S79], and J3DPerfUnit, which was used to support performance testing in ASD [S114]. Feitelson et al. [S53] report Perflab, a tool that help examining the effect of adding a new code on the performance of system prior to pushing new code on production servers, in the context of continuous deployment. Moreover, the authors report using Gatekeeper tool to conduct dark launch of codes on different servers and test QRs such as scalability and performance.

#### *4.2.7 Guidelines*

Seven primary studies reported guidelines for managing QRs in ASD and RSD. Two of the studies report guidelines related to QRs in general [S132, S156], two [S12, S127] on security, two [S81, S140] on usability, and one [S38] provides guidelines addressing availability, scalability, and performance. Specifically, these studies' guidelines relate to the integration security in ASD [S12], usability-pattern requirement analysis for managing usability in ASD [S81], and catalog of DevOps patterns that have been proven to enhance the availability, scalability, and performance of web applications based on experiences [S38]. For instance, Barbosa et al. [S12] provide guidelines for integrating security techniques into agile projects without undermining the agility of the process. The guidelines provides recommendations, which include security awareness training, introducing a security master role, creating a security backlog, and incorporating evil stories to detect vulnerabilities, which are presented with descriptions and the presumed benefits based on literature and expert interviews.

### 4.3 Challenges of managing QRs in ASD and RSD (RQ3)

We found 102 primary studies reporting the challenges of managing QRs in ASD and RSD. Out of these only three primary study [S96, S137, S144] reported challenges in RSD context. We applied thematic analysis [60] to synthesize the challenges and found 22 challenges, 18 of which were recurring (i.e., they were reported in more than one primary study; see Table 3). The limited ability of ASD to handle QRs is the most commonly reported type of challenge, and time constraints due to short iteration cycles was the second most common.

**Table 3 Frequency of reported challenges (studies reporting challenges in RSD context are in bold fonts)**

| **Challenge** | **Primary studies** | **Frequency** |
|---|---|---|
| Limited ability of ASD to handle QRs (e.g. eliciting, reasoning, modeling, and planning QRs) | S2, S9, S30, S42, S45, S49, S50, S51, S52, S60, S67, S75, S85, S90, S94, S109, S115, S116, S135, S155 | 20 |
| Time constraints due to short iteration cycles | S9, S12, S13, S20, S23, S28, S30, S60, S85, S117, **S137**, S140, S148 | 13 |
| Limitations in testing QRs | S13, S16, S20, S31, S37, S41, S72, S107, S122, S129, S130, S155 | 12 |
| Neglect of QRs | S2, S3, S8, S12, S19, S33, S54, S80, S105, S107, S129, S150 | 12 |
| Lack of an overall picture of QRs | S23, S30, S31, S44, S129, S130, S139 | 7 |
| Overlooking of QRs by customers | S31, S67, S90, S107, S115, S121, S129 | 7 |
| Customers' lack of awareness of QRs | S2, S15, S67, S75, S121, S122 | 6 |
| Limited QR expertise in ASD teams | S12, S31,S92, S116, S121, S122 | 6 |
| Prioritizing only business value | S19, S33, S103, S107, S155 | 5 |
| Late consideration of QRs | S3, S8, S130, S155 | 4 |
| QR documentation challenge | S15, S132, S138, S155 | 4 |
| Budget limitations | S12, S31, S86 | 3 |
| Communication challenges | S22, S129, S130 | 3 |
| Unclear responsibilities regarding management of QRs | S1, S15, S31 | 3 |
| Challenges with verification of QRs | S92, S122 | 2 |
| Unclear QRs | S1, S31 | 2 |
| Overlooking sources of QRs | S129, S130 | 2 |
| Software architecture challenge | S129, S130 | 2 |
| Absence of security and privacy experts in feature implementation discussions in RSD | **S144** | 1 |
| Difficulty of splitting QRs into user stories and describing their interactions as they can be linked to many user stories | S109 | 1 |
| Forgotten usability goals (i.e., usability goals written in user stories are not used in the design phase or while evaluating user stories) | S30 | 1 |
| Slow feedback loop for meeting QR targets | **S96** | 1 |

*The limited ability of ASD to handle QRs* due to insufficiencies related to eliciting, analyzing, modeling, documenting, and managing QRs was reported in 20 primary studies. Nine studies [S2, S30, S42, S45, S60, S67, S75, S90, S115] reported challenges related to eliciting and analyzing QRs, two [S51, S52] reported that ASD was insufficient for identifying activities to plan and visualize QRs, one [S155] reported insufficient planning of QRs, and two [S49, S50] reported that it lacks techniques for eliciting, modeling, and linking QRs with functional requirements in advance. User stories in ASD were also reported to be inadequate for specifying QRs [S30, S90, S94, S109, S116, S135].

*Time constraints due to short iteration cycles* is the second most commonly reported category, identified in 13 primary studies. For instance, Scrum teams may not address QRs such as security, completely and on time due to short iteration cycles [S10, S60, S117]. Additionally, existing ASD practices for engineering QRs are not suitable for short iteration cycles, which makes it challenging to integrate QRs without compromising time and cost [S12]. For instance, practices for managing security QRs are heavyweight to fit well in short sprint iterations [S13, S23]. Time constraints also minimize the focus on addressing QRs such as usability [S28, S30]. Additionally, in RSD, where software can be deployed several times a day, handling QRs like security is difficult [S137].

*Limitations in testing QRs* are the third most commonly cited category of challenges. In ASD, systematic approaches for testing QRs are lacking, and writing test cases and defining an acceptance test for QRs is problematic [S37, S41, S72]. For instance, ASD teams may lack formal procedures and guidelines for testing specific QRs (e.g., stability, security, and usability) [S37, S72, S107]. Isomursu et al. [S72] identify the lack of clear steps for planning and executing usability testing as a common challenge in ASD projects. QRs are hard to test due to their cross-functionality and the difficulty of clarifying them for customers [S31]. Additionally, agile testing practices' inadequate focus on QRs, combined with the complex functionalities such as user interaction, may exacerbate the challenges of managing them in ASD [S20].

*Neglect of QRs* is tied to the third most frequently reported category of challenges with limitations in testing QRs. For varying reasons, ASD teams often pay little attention to or ignore QRs. They may emphasize implementation of functional features and neglect QRs [S8, S54, and S105]. Sometimes, practitioners may be reluctant to implement or address QR issues until this is explicitly needed for visible features [S80]. In cases in which software is developed and deployed internally, agile teams report that they ignore QRs such as security. These teams ignored security QRs by relying on the infrastructure and control mechanisms of their organization set in place to protect the environment [S3]. Agile teams neglecting QRs may experience increased costs [S80] and performance issues [S19] in later stages of development, incurring technical debt.

*Lack of an overall picture of QRs* is a category that was reported in seven primary studies. It includes challenges in ASD that arise from a lack of comprehensive understanding of QRs. For instance, ASD fails short in determining broad picture of implementing security requirements [S23], and ASD teams may find it difficult to maintain a comprehensive vision of QRs such as usability [S30]. The difficulty of defining and maintaining an overall picture of QRs causes agile teams to face ambiguity and challenges regarding quality issues [S31].

*Overlooking of QRs by customers* refers to ASD customers' failure to acknowledge the importance of QRs. Agile customers do not recognize the significance or impact of QRs [S31, S67, S107, S115, S121] or their business value [S90]. This is especially common in the early stages of ASD [S107]. While failing to acknowledge the value of QRs, customers also may not allocate an adequate budget for QR-related tasks. They may only realize the importance of QRs when they face issues related to performance or security in later stages [S31].

*Customers' lack of awareness of QR*s is another category of challenges. Very often, agile customers do not have adequate knowledge to specify and communicate their needs regarding QRs [S2, S67, S75, S121, S122]. For instance, customers may not consider the internal qualities of a system, such as maintainability, portability, and reusability [S2, S67]. They may find specifying certain QRs, such as security, difficult if they lack knowledge about the QR [S121, S122].

*Limited QR expertise within ASD teams* is another category of challenge reported in six primary studies [S12, S31, S92, S116, S121, S122]. Inexperienced team members who lack QR skills may emphasize implementation of functional requirements and ignore QRs [S31, S122]. Additionally, team members such as the product owner may not have all the required skills to implement specific QRs (e.g., user interaction, security). Finding professionals with expertise in QRs such as security is also challenging [S12].

*Prioritizing only business value* when it comes to requirement prioritization and development goals results in challenges of managing QRs in ASD [S19, S33, S103, S107, S155]. ASD management teams often fail to consider QRs such as security and choose to prioritize feature development goals [S103]. However, when ASD teams focus on prioritizing only business value, they may fail to address QRs and face delays in implementation. For instance, Cao et al. [S33] reported that an ASD team that focused on business value prioritization faced challenges in system security and efficiency, which affected the success of the project.

*Late consideration of QRs* is another category of challenges related to managing QRs in ASD. It was reported in three primary studies [S3, S8, S96]. Handling security issues in late phases of development may affect agile projects because it requires introducing many changes [S8]. Nilsson et al. [S96] also revealed that late testing of QRs results in performance degradation and unpredictable effort estimation.

*QR documentation challenge* can arise when QRs are not properly documented. For instance, outdated, unclear or missing documentation of QRs [S15, S132, S155]. Additionally, handling QRs such as safety and security is challenging in ASD, as they require end to end documentation which is enforced by different regulations [S155].

*Budget limitations* in terms of cost and time lead to challenges when managing QRs in ASD. Three primary studies reported difficulty of managing QRs due to the lack of budget allocated to QRs [S12, S31, S86]. For example, Camacho et al. [S103] show how budget limitations affect the implementation of security requirements in ASD. According to the authors, ASD customers fail to allocate a budget for security because they are not aware of its value [S31].

*Communication challenges* occur when there is cognitive gap regarding QRs among agile teams or in the presence of hidden assumption regarding QRs. Additionally, when QRs are not visible they lead to difficulties in QR communication process [S129, S130]. Limited domain knowledge regarding QRs among agile teams and their members resulted in misinterpretations of QRs [S22].

*Unclear responsibilities regarding management of QRs* in ASD is a category that was cited in three primary studies [S1, S30, S72]. ASD teams fail to address QR aspects by failing to assign individuals the responsibility of handling QRs. Specifically, two studies [S30, S72] revealed the challenges related to management of user experience that resulted from a lack of clear usability responsibility in ASD teams.

We also found two primary studies [S92, S122] that reported *verification of QRs* as a challenge, as it is difficult to define verification scenarios for QRs. Two different primary studies [S1, S15] also reported that *unclear QRs* are challenges, revealing how unclear security requirements can cause problems in ASD. *Overlooking of sources of QRs,* and *Software architecture challenges* due to unmanaged QR architectural changes and suboptimal priorities of conflicting QRs in ASD, are other challenges reported in two primary studies [S129, S130].

We also found four non-recurring challenges of managing QRs: *difficulty of splitting QRs into user stories and describing their interaction* as they can be linked to many user stories [S109]; *forgotten usability goals (i.e., usability* *goals written in user stories are not used in the design phase or while evaluating user stories)* [S30]; *absence of security and privacy experts in feature implementation discussions in the context of RSD* [S144] and a *slow feedback loop for meeting QR targets* in the context of embedded ASD employing continuous integration [S96].

### 4.4 Mapping QR management strategies to challenges

In order to see which challenges were the focus of management strategies and highlight the challenges that require more attention, we mapped the QR management strategies to the challenges. The actual mapping is available here. Fig. 5 shows the mapping of the QR management strategies that address a specific challenges.

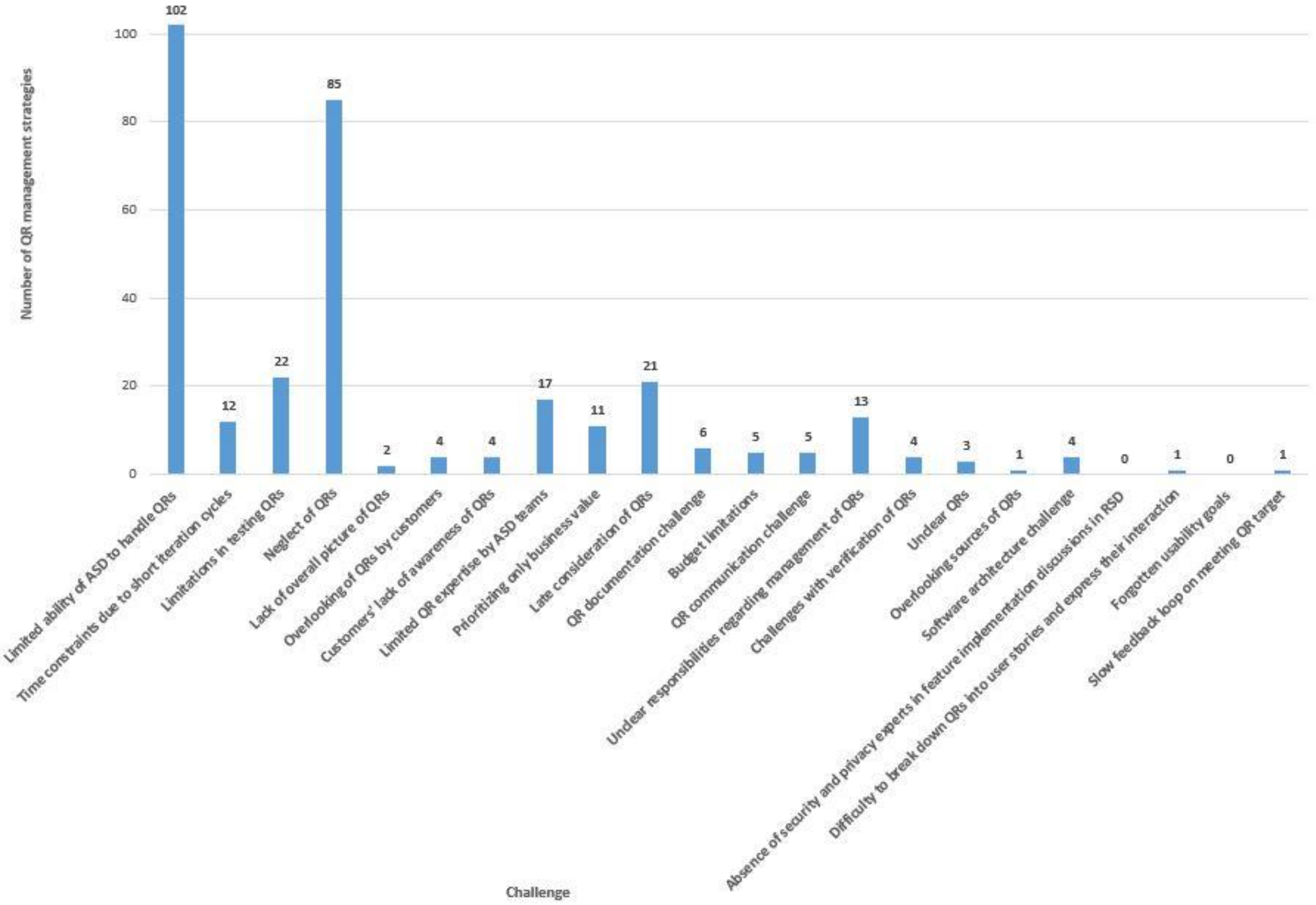

**Fig. 5. Mapping QR management strategies to challenges**

As shown in the figure, most strategies tend to address two challenges: the limited ability of ASD to handle QRs and neglect of QRs. However, many management strategies do not address another commonly recurring challenge, time constraints due to the short iteration cycles of ASD.

## 5. Discussion

In this section, we discuss the implications of our findings for SE research, SE industry, and SE education, and compare our findings with related work.

### 5.1 Implications for SE research

We identify the implications for SE research by analyzing the overview of research, strategies for managing QRs, challenges, and mapping of strategies to challenges.

#### 5.1.1 Overview of research

While there is a wide body of literature on the management of QRs in ASD, the stream of literature investigating the management of QRs in RSD is still nascent. However, there is increasing adoption of RSD approaches in which ensuring quality is important, which demands new skillsets regarding QRs (e.g., security-related skills and QR-testing skills) [1]. Therefore, SE research should focus on identifying the challenges and opportunities for improving the management of QRs as well as ways to ensure software quality in RSD.

Our results revealed that a large number of empirical studies (106) have been conducted on this topic, in contrast to previous secondary studies [10,17,22], which reported a smaller number of empirical studies. For instance, Alsaqaf et al. [9] reported that only 29 of the 60 primary studies they identified were empirical in nature. Villamizar et al. [22], reported 21 empirical studies. One reason for this difference may be the snowballing search strategy we applied, which revealed a relatively large number of primary studies. Additionally, the search period in our SMS included primary studies published until July 2019. While the trend of conducting empirical research seems promising, the quality assessment results (see Fig. 4) reveal that the literature on management of QRs in ASD and RSD needs to improve its reporting of research rigor. Therefore, researchers should pay enough attention to the reporting of aspects related to research rigor (i.e., study design, context, and validity) and not focus only on industrial relevance aspects. Research reports that provide detailed information about the industrial relevance and research rigor ease the transferability of knowledge and are more beneficial to the SE community [56]. This is exemplified by the 16 primary studies with high industrial relevance (4) and research rigor (3) scores in our study, which provide good insight into the management of QRs in ASD and RSD.

#### 5.1.2 Strategies for managing QRs

While primary studies have empirically evaluated strategies for managing QRs in real-life settings through case studies, action research, surveys, and experiments, the evaluations are usually limited to a single scenario. Extending evaluations to wider and more varied contexts is important to increase their generalizability.

Regarding QR management strategies in ASD, our results show that security, usability, and performance QRs receive more attention. In the primary studies reporting strategies for managing QRs in RSD, deployability, performance and availability received more attention. Security was addressed by three of the 13 primary studies in RSD context [S106, 137, S142]. Although this result does not imply that security receives less attention in RSD, we believe that future studies should not overlook security, as it plays important role in RSD [62].

Regarding security QRs, although there are proposals that have been empirically evaluated and found effective in handling aspects of security in ASD (e.g. EAST method, TDD with quality attribute focus, NORMATIC), it is important to extend the empirical evaluations to other contexts and also ensure that these strategies cover wide aspects of security.

Regarding the distribution of QR management strategies, we observed that there were more practices and methods than models, frameworks, advice, tools, and guidelines. Our results demonstrate the lack of tools and guidelines for managing QRs, which is also reported in some studies [63,64]. Thus, future research should focus on providing more tools and guidelines (e.g., guidelines for testing QRs, tools supporting specification and prioritization of QRs) to aid management of QRs in ASD and RSD. Additionally, we observed that most proposed models and frameworks lacked supporting tools and guidelines. Model and framework QR management proposals would be more beneficial when they are accompanied with tools and guidelines.

#### 5.1.3 Challenges of managing QRs

Interestingly, the challenges due to the lack of an overall picture of QRs, unclear QRs, and QR documentation challenge in ASD are related to technical debt. For example, Soares et al. [65] identify incomplete documentation and incomplete design specification as indicators of documentation debt, and Behutiye et al. [4] identified a lack of understanding of the system being built (functional requirements and QRs) as a significant cause of technical debt in ASD. Detailed investigation of the relationship between technical debt and QRs may provide insight into better ways to manage QRs.

Regarding the challenges of managing QRs in RSD, we observed two specific challenges, *absence of security and privacy experts in feature implementation discussions in RSD* and *slow feedback in meeting QR targets.* The third challenge which is *time constraint due to short iteration cycles was* also common to ASD contexts.

#### 5.1.4 Mapping QR management strategies to challenges

Time constraints due to short iteration cycles received little attention by the QR management strategies, despite the fact that it being the second most common challenge (see Fig. 5). We infer the importance of lightweight strategies for managing QRs, as these are suitable for short iteration cycles. Strategies that propose additional roles and artifacts should not compromise the rapidness of the process. Therefore, we suggest that future proposals for managing QRs consider time constraint aspects and be lightweight enough to support short iteration cycles.

Although the limitations associated with testing QRs include deficiency in testing practices, difficulty of testing QRs (e.g. specifying acceptance tests, testing the cross-functional aspects of QRs), and lack of guidelines for testing QRs, the strategies we found for solving these challenges were mainly focused on testing specific QRs (e.g., performance, security, and usability). We did not find strategies that clearly show how QR testing can consider the cross-functional aspects of QRs in ASD. We also did not find guidelines that support testing of QRs, even though it was reported as a limitation in testing QRs in ASD. Future research should address this gap, as guidelines for testing and managing QRs may help in the software development process.

The lack of an overall picture of QRs, customers' lack of awareness and overlooking of QRs are also significant challenges that received little attention in strategies. Absence of security and privacy experts in feature implementations discussions in RSD, and forgotten usability goals received no attention in the QR management strategies. Future research and proposals on QR management strategies should consider addressing aspects reflected in these challenges too.

### 5.2 Implications for the SE industry

Our SMS provides a list and classification of QR challenges and management strategies to help industrial practitioners identify and manage QR challenges in ASD and RSD. From empirical findings reported in our SMS, project managers, agile development team (developers, product owner, tester and scrum master) and other stakeholders (e.g. customers, usability designers) can learn about potential challenges of management of QRs in ASD and RSD, and hence be able to take a more systematic and preventive approach from incurring similar challenges. For instance, excluding security and privacy experts from discussions regarding implication of features under development, which developers and operations people hold iteratively in RSD context, may increase the risk of security not being addressed well in RSD. Project managers can minimize such risk by including security and privacy experts while having similar discussions in their context. ASD teams and project managers can learn importance of assigning clear responsibilities regarding QRs (e.g. usability) in order to prevent challenges from *the lack of clear responsibility regarding QRs* in ASD [S30, S72]. Regarding the challenges from *limited expertise on QRs*, product owners and developers can learn the importance of building skills on QR, and hence educate themselves on QRs. Project managers and developers can also recognize that the *neglect of QRs* can be caused when developers rely on built in infrastructure, or become reluctant from implementing QRs and hence take proactive approach and preventive actions, e.g. by preparing guidelines with procedures for handling QRs that development team should follow.

We also found important strategies that ASD product owners and developers can use such as EAST [S46], and SEAP [S23] to address security, NERV methodology [S42] which was effective in eliciting and documenting QRs in ASD. ASD and RSD teams can identify that investing in education on QRs (e.g. training on security and usability) is helpful in addressing QR challenges. Additionally, they may introduce specific roles for QRs (e.g. security experts to handle security and privacy issues, and usability experts to help with usability issues). Similarly, product owners can adopt practices such as specifying QRs in Definition of ready of user stories, to manage QRs better. Our SMS is also informant for testers about tools such as SQUISH, which has been used for automating usability testing in ASD, or Perflab that helped in monitoring performance in RSD context.

We observed that the management strategies often introduce an additional role or team that is mainly responsible for addressing QRs. For instance, these strategies assign security experts [S7], security masters [S9], delivery story teams [S39], architecture groups [S40], liaison officers [S69], and usability experts [S113, S124]. These practices may be helpful for addressing the challenge of unclear responsibilities regarding QRs in ASD. However, it is also important to ensure that additional roles do not compromise the agility of the software development process.

While considering the neglect of QRs, we noticed that ASD teams ignore QRs for reasons other than an emphasis on implementation of functional requirements. For instances, teams may rely on existing internal infrastructure or wait until there is a request related to the QRs or they have a significant impact [S3, S80]. This suggests that it is important to increase agile teams' awareness of QRs and their impact on the software development process. It may also be helpful to create roles that are responsible for guiding the implementation of QR within ASD teams. Barbosa et al. [S12] show how a security master, who is responsible for educating agile teams about security requirements and implementation, guiding security-related task planning, and ensuring security objectives are met, can be helpful for instilling a culture of addressing security concerns in agile teams, thus minimizing neglect of QRs.

While dealing with QRs, it is important to assess the combined effect of challenges. For instance, a challenge related to customers' lack of knowledge of QRs could be exacerbated if inexperienced ASD teams are working with the customer and there is a cognitive gap regarding the communication of QRs among members of the ASD team.

Most QR management strategies do not address overlooking of QRs and customers' lack of awareness, which are frequently reported challenges in our study (see Fig. 5). Generally, ASD promotes customer collaboration [7]. In this regard, ASD teams should help customers deal with QRs. For instance, experienced developers may guide customers while eliciting QRs and making decisions about QRs in the early phases of development. Customers should also educate themselves about QRs to ensure clear communication of their QR needs with other stakeholders. Furthermore, clear and early consideration of QRs may benefit agile projects by allowing teams to avoid redundant work and additional costs [S75].

Regarding the limited QR expertise in ASD teams, some strategies aim to address skillset-related concerns by improving ASD team members' and product owners' QR skills. However, we noticed that this challenge is also relevant in RSD contexts, where technology increases the need to learn new skill sets in order to manage quality. We highlight the need for building quality engineering skills to overcome challenges related to limited expertise in QRs, similar to the latest world quality report [1].

Although we found strategies for handling security in ASD and RSD, security remains being a concern. One reason can be the difficulty in meeting all aspects of security. It is also possible that rapid release cycles in ASD and RSD are problematic for QRs such as security. Thus, security receives more attention than other QRs. We observe similar trends in a recent world quality report by Capgnemini [1], where enhancing security is identified as the highest objective of IT strategy by respondents from ASD and RSD companies.

We observed that privacy requirements are also investigated less compared to security in ASD. The primary studies we found address privacy together with other QRs, e.g., AFFINE framework, which enabled earlier consideration of QRs (privacy, security, and usability mainly) [S26, S27], and threat poker method for analyzing security and privacy risks in ASD [S149]. A potential reason that privacy is under investigated in ASD and RSD, can be that activities related to privacy are perceived and treated separately from software development [S144]. Regarding, demands of regulations such as GDPR (General Data Protection regulation) on IT systems [S149], and the limited research investigating privacy in ASD and RSD, it is important that SE industry recognize the significance of privacy issues. Moreover, ASD and RSD teams and management should invest and acquire skills and resources on privacy issues. Safety is another QR that received less attention among the primary studies. One reason can be that safety is considered as silo activity from actual software development. On the other hand, it is possible that the standard regulations and conformances, which require strict documentation for ensuring safety [S138], are less feasible with ASD and RSD.

Finally, the results of classifying authors' affiliations (see Fig. 2 (a)) indicate the need for enhancing academia–industry collaborations and contributions from the industry. Academia–industry collaborations guarantee the applicability and impact of SE research [66]. Furthermore, they help identify research problems and propose solutions that are relevant to both the industry and academia.

### 5.3 Implications for SE education

Our findings suggest that SE education should incorporate topics on QRs, and specific ways of handling them in ASD and RSD contexts. For instance, students should be familiarized with product quality standards like ISO/IEC 25010, and be aware of the significance of QRs, and their impacts on quality and cost. SE education can contribute in addressing existing challenges (e.g. limited ability of ASD in handling QRs, limited expertise in QRs) by enabling students acquire the required skills. Exercise sessions in RE and SE courses can be tailored to incorporate different strategies on handling QRs in ASD and RSD contexts. It is also important that SE courses realize and meet the real needs of the SE industry. For instance, modern RSD approaches require skills in continuous integration, automated testing and deployment [S144]. Our findings also showed a need for addressing the limited expertise on QRs such as security both in ASD and RSD context. In this regard, SE education should be up to date with the demands of the SE industry.

### 5.4 Related work

Alsaqaf et al. [10] investigated agile practices related to engineering QRs in large-scale ASD projects and solutions for engineering QRs in ASD. They report practices of face-to-face communication, delivery stories, test-driven development (TDD), and iterative emergence of requirements, frequent requirement prioritization, and pair programming. Our results reveal similar practices such as clear discussion and communication about QRs among customers and agile team members [S31], delivery stories [S39], TDD focused on quality attributes [S20], and incremental and iterative approaches (e.g. the incremental architecture approach [S63]). Schön et al. [11] identified management strategies such as the AFFINE framework for integrating QRs [S26, S27] and NORMAP [S48] and SENOR for eliciting QRs [S94], which are also included in our findings. We also found practices that use the definition of ready of user stories [S104], practices that use artifacts (e.g., UserXstories to specify interactive usability stories in ASD [S36], shadow backlogs for usability [S125], abuser stories to describe security requirements [S101, S143], and misuse stories for security [S78, S143]). Moreover, our findings reveal practices for managing QRs in RSD, such as experimentation with A/B testing [S85], automated testing, and monitoring and deployment of system security [S106], and framework to assist architecting continuous delivery by introducing QR and operational aspects [S152], which were not discussed in previous secondary studies.

Villamizar et al. [22] identified approaches for handling security in ASD that introduce artifacts such as abuser story, present approach for integrating security in Scrum, methods such as NORMAP [S48], and NERV [S42], and NORMATIC tool [S48, S50], and guidelines for eliciting and documenting QRs [S132, 156], that are included in our findings. They also reported a way for mapping security activities into Dynamic System Development Method (DSDM) [67] that was not found in our study. We also found strategies such as Secure Feature Driven Development (SFDD) [S54], and CIViT [S96] models, method to integrate security in feature driven development [S58], and tagging approach to elicit security in ASD [S121], and practices of security oriented TDD [S95] that were not reported in their study.

Silva et al. [48] identified artifacts in ASD, such as user stories, scenarios, personas, low-fidelity prototypes and practices for close collaboration, little design upfront, and the need for a big picture to deal with usability in ASD. We found similar practices (e.g., those utilizing scenarios and low-fidelity prototypes) that were oriented towards QRs in general in addition to usability in particular. Our findings also reveal methods such as usability pattern-based analysis, InterMod, and U-Scrum; practices that apply artifacts, such as UserXstories, shadow backlogs, introduction of usability expert roles; models such as the agile usability software model and US-Scrum; tools such as POLVO and SQUISH; frameworks such as UCD; and guidelines and advice for developing clear usability goals.

We found 18 categories of challenges and four non-recurring challenges. The limited ability of ASD to handle QRs was the most popular category (see Table 3). This challenge relates to the insufficiency of user stories and the lack of techniques for identifying, eliciting, modeling, planning, and visualizing QRs in ASD. Inayat et al. [18] also reported the limitations of agile approaches in handling QRs. Alsaqaf et al. [10] reported similar challenges—inability of user stories to document QRs, and agile does not provide a widely accepted technique for gathering the QRs—as two separate categories. In our SMS, we found primary studies that focused on addressing these limitations. For instance, artifacts such as evil user stories (to address security) [S12], UserXstories (to create user interaction stories) [S36], the NFRusCOM card of the NERV method [S42], and abuser stories (to create security requirements) [S101] were used to overcome the inadequacy of user stories for specifying QRs in ASD.

Neglect of QRs, which is the third most common challenge in our SMS, was also reported in previous secondary studies [10,11,16–18,21]. We found that ASD teams ignore QRs for different reasons. For instance, teams focus on the delivery of functional features, wait until QR features and problems become visible, or completely rely on the infrastructure and environment of their organization to address specific QRs. However, it is important that ASD teams recognize the significance of QRs and handle them properly during the development process. We found several strategies that focus on overcoming this challenge, such as weaving quality attribute concern data into the software lifecycle [S18], prototyping with a focus on quality [S19], early consideration and discussion of QRs by teams and customers [S41], by helping ASD teams to focus on QRs.

Compared to Alsaqaf et al. [10], we found 11 new categories of challenges of managing QRs. These are time constraints due to short iterations, limitations of testing QRs, customers' lack of awareness of QRs, overlooking of QRs by customers, lack of an overall picture of QRs, prioritization of only business value, budget limitations, unclear responsibilities regarding management of QRs, QR documentation challenges, communication challenges, and unclear QRs. We also found three non-recurring challenges: absence of security and privacy experts in feature implementation discussions in RSD, slow feedback loop for meeting QR, and forgotten usability goals. On the other hand, the findings from Alsaqaf et al. [10] reveal 12 challenges, five of which were not included in our study. These include product owner's heavy workload, insufficient availability of product owner, insufficient requirement analysis, dependence on the product owner as a single point to collect requirements, and focusing on delivering functionality at the cost of architecture flexibility.

We found six challenges of managing QRs that were not reported in previous secondary studies: *time constraint due to short iterations, lack of an overall picture of QRs, overlooking of QRs by customers, unclear responsibilities regarding management of QRs, unclear QRs, and absence of security and privacy experts in feature implementation discussions in RSD*. We also found customers' lack of awareness of QRs, prioritization of only business value, budget limitations, QR documentation challenges, and QR communication challenges. However, we observed that other secondary studies [11,16–18] have reported closely related findings as general challenges of RE in ASD. These challenges include customers' insufficient knowledge on domain, challenges from requirements documentation, communication, prioritization, budget limitations and lack of information.

Security, performance, usability, maintainability, and reliability are the top five most frequently reported QRs in our study. Similarly, Zou et al. [68] found that usability and reliability are the most commonly discussed types of QRs among developers after analyzing stack overflow data (i.e., posts and comments related to QRs). The authors also identified that maintainability and efficiency received less emphasis among developers in their study. While our findings regarding usability, maintainability, and reliability are in line with the findings of Zou et al. [68], security, which was the most frequently reported type of QR in our study, received no attention among the developers in their study. One possible reason for this disparity is the significance of security in ASD and RSD [1]. In addition, security requirements may be more difficult to discuss, and developers may lack skills concerning security-related issues [S31], thus leading to less discussion of security among developers.

## 6. Conclusion

Due to advancements in technology and highly competitive markets, ASD and RSD are a necessity for fast delivery of software. Ensuring high software quality is greatly important in such development approaches. QRs affect software quality and have economic implications on software development, and they play an important role in determining the success of ASD and RSD projects.

In this study, we explored the management of QRs in ASD and RSD through classification and synthesis of 156 primary studies. Our findings revealed the extensive work carried out in primary studies, affirming the significance and relevance of the topic for both researchers and the industry. The domain classification reinforces the significance of QRs in various domains, especially technology and industrial goods and services.

We found that 68% (106 of the 156 primary studies) were empirical studies, while 10% (16) were experience reports. This is an encouraging finding, as the knowledge produced by these primary studies is based on investigation of QRs in real-life contexts and the lessons learnt while dealing with QRs.

Regarding QR type, security was the primary focus of many primary studies investigating the management of QRs in ASD. However, deployability, performance and availability received a higher priority than security among the 13 primary studies investigating the management of QRs in RSD.

We found 143 primary studies proposing QR management strategies, 93 of which were empirical. We identified a comprehensive list of strategies, including 74 practices, 43 methods, 13 models, 12 frameworks, 11 advices, 10 tools, and 7 guidelines. The QR management strategies mainly focused on addressing the limited ability of ASD to handle QRs and neglect of QRs.

We also found 18 categories and 4 non-recurring challenges of managing QRs. The limited ability of ASD to handle QRs, time constraints due to short iteration cycles, limitations in testing QRs, and neglect of QRs were the most significant challenges reported in the literature. We also found challenges such as unclear QRs, QR documentation challenges and lack of an overall picture of QRs, which are indicators of technical debt.

For researchers, our work identifies important research gaps, such as the need for more tools and guidelines supporting management of QRs, improvement in the reporting of research rigor aspects by empirical studies, and investigation of the link between QRs and technical debt in ASD and RSD. Additionally, time constraints due to short iteration cycles, limitations regarding testing of QRs, lack of awareness and overlooking of QRs by customers are significant challenges that need to be focused upon by future research proposing QR management strategies in ASD and RSD. Future QRs management strategies aiming to address the limitations of testing QRs, should also consider cross-functional aspects of QRs. Security remains being a concern in ASD and RSD, and researchers should give enough attention towards addressing it. Our findings also indicate that, despite the various evaluations and applications of QR management strategies in specific contexts, they have not been validated in varying or wider contexts. Therefore, we encourage that future studies do this to increase the generalizability of their findings.

For the SE industry, our work synthesizes QR management strategies and challenges and maps strategies to challenges, which may be beneficial for practitioners. Additionally, our study shows the need for increasing the contributions of the SE industry and academia–industry collaborations on this topic.

For SE education, our work reveals the need for improving SE and RE education by covering aspects of QRs in ASD and RSD. Moreover, it shows the need for tailoring these courses to meet up to date needs of the SE industry.

**Acknowledgments.** This work is the result of the Q-Rapids project, which has received funding from the European Union's Horizon 2020 research and innovation program under grant agreement no. 732253.

## Appendix A. Data properties

D1.1 *Publication year*: The year of publication.

D1.2 *Publication venue*: The place that the study was published (i.e., conference, workshop, journal, book chapter).

D1.3 *Type of research (empirical/theoretical/experience report)*:

- Empirical: A study that bases its findings on direct empirical evidence. The following empirical research methods were reported by the authors of the primary studies:
  - Case study: A case study or exploratory study in which the researchers analyze and answer predefined questions for one or multiple cases.
  - Survey: A study that administers a survey through questionnaires, observations, or interviews.
  - Action research: A study that applies a research idea in practice and evaluates the results (i.e., a cross between an experiment and a case study).
  - Experiment: A study that empirically investigates causal relations and processes.
  - Statistics analysis: A study that performs quantitative analysis on real-life project data.
  - Mixed method: A study that uses several research methods.
  - In addition, empirical studies applying interviews, and observations.
- Theoretical: A study that is based on an understanding of a certain field but lacks empirical evidence to support its findings or suggestions.
- Experience report: A study that reports the lessons learned. Unlike empirical studies, experience reports usually do not present the details on study design, validity and research questions.

D1.4 *Reported domain*: The domain (e.g., telecommunication, automotive) that is the focus of the study.

D1.5 *QR type*: The type of QR that is explicitly stated in the primary study (e.g., usability, security).

D1.6 *Research rigor*: The research method of an empirical study. Following Ivarsson and Gorschek [56], the context, study design, and validity are scored as weak (0), medium (0.5), or strong (1).

- Context (e.g., development mode, speed, company maturity)
  - A strong description (1) describes the context to a degree that the reader can understand and compare it to another context. This involves descriptions of the development mode (e.g., market-driven), development speed (e.g., short time to market), or company maturity (e.g., start-up, market leader).
  - A medium description (0.5) briefly mentions or presents the context in which the study is performed, but not to a degree that a reader can understand and compare it to another context.
  - A weak description (0) has no description of the context in which the evaluation was performed.
- Study design/research method (i.e., measured variables, treatments, controls used in the study)
  - A strong description (1) describes the study design to a degree that a reader can understand the variables measured, the control used, the treatment(s), the selection/sampling method used, and so on.
  - A medium description (0.5) briefly describes the study design (e.g., “six researchers performed steps 1, 2 and 3”).
  - A weak description (0) does not describe the study design of the presented evaluation.
- Validity (i.e., threats to validity, measures/mitigations to limit threats)
  - A strong description (1) discusses the validity of the evaluation in detail, describing threats and providing details regarding measures to limit them. Also, different types of threats to validity (e.g., conclusion, internal, external, and construct) are described.
  - A medium description (0.5) mentions the validity of the study, but provides no details.
  - A weak description (0) does not describe any threats to the validity of the evaluation.
- Total research rigor value (i.e., the sum of the context, study design, and validity values)

D1.7 *Industrial relevance*: Includes aspects such as subjects, context, scale, and research method. Following Ivarsson and Gorschek [56], industrial relevance is scored as: “contributes to relevance” (1) and “does not contribute to relevance” (0).

- Evaluation of the realism of the study environment
  - Subjects (practitioners, students, researchers)
    - The subjects contribute to relevance (1) if they represent the intended users of the technology (i.e., industrial professionals).

- The subjects do not contribute to relevance (0) if they are not representatives of the intended users of the technology (e.g., students) or the study does not mention subjects.
  - Context (industrial setting)
    - The evaluation contributes to relevance (1) if it is performed in a setting that represents the intended usage (i.e., an industrial setting).
    - The evaluation does not contribute to relevance (0) if it occurs in a laboratory or a setting that does not represent a real usage situation.
  - Scale (realistic size, usefulness, scalability)
    - The scale of the application contributes to relevance (1) if it is of realistic size (i.e., industrial scale).
    - The scale of the application does not contribute to relevance (0) if it is of unrealistic size (e.g., toy examples).
- Evaluation of the research method
    - The research method contributes to relevance (1) if it facilitates investigating real situations, is relevant to practitioners, and explicitly stated. Research methods that contribute to relevance may include case studies, surveys, interviews, controlled experiments, and action research.
    - The research method does not contribute to relevance (0) if it does not investigate a real situation (e.g., conceptual analysis, laboratory experiment, software experiment).
- Total industrial relevance (i.e., the sum of the subjects, context, scale, and research method values)

D1.8 *Authors' affiliation*: Refers to whether the authors of the paper are from academia, the industry, or both (i.e., academia–industry collaboration).

D2. *Strategy to manage QRs*: Explicitly stated advice, frameworks, guidelines, methods, models, practices, and tools for managing QRs in ASD and RSD. We adapted the classification used in [58,59] and introduced definitions of practices and advice to classify strategies that were not explicitly reported in the aforementioned classification. Thus, our classification includes the following:

- Advice: recommendations and suggestions for managing QRs
- Frameworks: conceptual maps for managing QRs
- Guidelines: systematically synthesized list of advice and recommendations for managing QRs
- Method: approaches and procedures for managing QRs
- Models: representation of an observed reality to help manage QRs
- Practices: applications of activities, ideas, and artefacts for managing QRs
- Tools: applications and technologies for managing QRs

D3 *Challenges*: Challenges reported in literature regarding management of QRs in ASD and RSD.

## Appendix B. Quality requirement management strategies

**NB** Primary studies marked with * report QR management strategies in RSD. Practices, methods, models, frameworks, and tools occurring in more than one primary study are **bolded**. Closely related practices (e.g., initial education on security requirements [**S7]** and security training [**S58**]) are also **bolded**.

| Primary study | QR management strategy | Category of the strategy |
|---|---|---|
| S7 | Practices for addressing security in ASD (1.**Initial education on requirements**, 2.Identifying and enumerating Security Requirements, 3.Agree on Definitions on requirements, 4.Applying Quality Gates (SDL) , 5.Secure Design Principles, 6.Counter Measure Graphs during design phase, 7. Applying Vulnerability & Penetration and security testing in testing phase 8. **Assigning security experts**) | Practices |
| S8 | Practice for establishing security requirements in XP via partitioning operations, and technical features of a system in the Planning Game practice | |
| S9 | Practices to integrate security principles in Scrum, 1.**Security backlog** used for security features analysis and implementation, 2.**Security Master role** managing security backlog in Scrum | |
| S15 | 1. Focus on adequate customer involvement, 2.focus on developers' awareness and expertise of security, and continuously improving the development process for managing security in ASD | |
| S17* | 1. Automated database scripts to minimize update time during database release, configuration tools to enable automatic deployment, 2. Clear understanding of stakeholder priorities and visibility of tradeoffs to make the right decisions for short and long-term deployability. | |
| S18 | Weaving quality attribute concerns into the life cycle of ASD | |
| S19 | Prototyping with quality focus to integrate architecture practices in into Scrum. | |
| S20 | 1. Release planning with architecture consideration, 2. **Prototyping with quality attribute focus**, 3.**Test driven development with quality attribute focus** and 4. architectural change that promotes stability in rapid fielding | |
| S31 | 1. Discuss non-functional aspects during project inception, sprint planning and user story development, 2.Team members reviewing | |
| | QRs and testing needs, 3.Clear communication between team members and customer, 4.Work with a quality mindset | |
| S32 | 1. Automated and continuous strategy to test robustness and performance; 2. Involving the product owner while defining what the success criteria for robustness and performance | |
| S35 | Iterative security architecture practice | |
| S36 | Apply UserX Stories to write interactive stories for usability in ASD | |
| S37 | Performance tests to address performance issues in ASD | |
| S39 | Practice of prioritizing QRs (establishing delivery story teams where delivery teams are defining architecture that meets all QRs, capturing QRs in user stories, use of delivery stories) | |
| S40 | Establishing architecture group responsible for addressing architectural management, development quality, and technical debt in ASD | |
| S41 | Practices for considering QRs in ASD (1. RE practices that consider NFRs in ASD feasibility study to determine system scope, 2.Early discussion on QRs between customers and agile team leaders, 3. Requirements elicitation through interviews, brainstorming, ethnography and requirements documentation ) | |
| S56 | 1. Applying state machines to design safety critical systems in ASD, 2. Applying continuous automated unit testing and 100% code coverage to ensure software quality in ASD of safety critical systems | |
| S58 | 1. **Security training** and 2. Defining fundamental security architecture before iterations commence to manage security in XP | |
| S63 | Incremental agile approach to architecture to address tradeoffs between performance, availability, security, and usability | |
| S69 | Liaison officer with requirements engineering role for addressing quality requirements and other stakeholder problems in ASD | |
| S78 | 1. Practices of misuse stories, automated testing of misuse stories and 2. Security review meetings after each iteration to address security in ASD | |
| S79 | Applying standard NFR tests where implementation is stopped and actions taken when deviations from ideal behavior is observed in ASD of embedded systems | |
| S80 | **Prototyping with a focus on reviewing big picture at steady interval and reviewing prototypes with a focus on QRs** | |
| S82 | 1.Quality scenarios to discover QR, 2.**Quality Attribute Workshop (QAW) durin**g XP's story production to elicit quality attribute requirements in the form of scenarios with the help of stakeholders | |
| S95 | **1.Security Backlog and 2.Security-oriented TDD** in ASD | |
| S98* | Experimentation through A/B testing to improve performance and customer satisfaction in continuous delivery | |
| S101 | Using abuser stories for describing security briefly and enabling security traceability in ASD | |
| S104 | Product owner specifies Definition of ready for user stories that contains architecture criteria on QRs | |
| S106***** | DevOps activities, such as automated monitoring, automated testing, and automated deployment of software can be helpful to a system's security | |
| S109 | Applying System Story to collect features related non-functional requirements that cannot be allocated in user stories | |
| S112 | Applying dedicated software architecture for addressing quality requirements in ASD | |
| S113 | Involving **Usability expert role** in ASD to address usability | |
| S119 | Discount usability (Scenarios, Card Sorting, Heuristic evaluation and thinking aloud) techniques in XP to address usability issues | |
| S122 | Practices for addressing security in Scrum (1.Reflective discovery of security needs, 2.Value and prioritization of security work, 3.Security expertise and security consultancy, 4.Verification and feedback with a focus on security) | |
| S124 | Practices to address usability in ASD (1. Low-fi prototypes, 2.Testing in between iterations with the application users, Usability designers and developers work in parallel, Usability designers should be involved in the project, 3.**Usability designers** should be fully integrated into the development team, 4.End users should be involved in the project, 5.**Employ workshops to introduce usability work**) | |
| S125 | 1.Applying shadow backlog to address usability, 2.organizing training sessions on usability and concretizing usability designers' knowledge on new systems | |
| S128 | Stress tests to determine the stability of the system and guarantee robustness, availability and no catastrophic crashing under heavy loads | |
| S129 | **1. Maintaining an assumption wiki-page 2.Use multiple product backlogs to include requirements of different viewpoints 3.Use automated monitoring tools 4. Reserve part of the sprint for important QRs 5.Sprint allocation based on multiple PBs 6.using preparation team to define QRs 7.Establish component team handling QR**s **of individual component 8.Establish QR specialist team 9.Use innovation and planning iteration** | |
| S130 | **1. Maintaining an assumption wiki-page 2.Use multiple product backlogs to include requirements of different viewpoints 3.Use automated monitoring tools 4. Reserve part of the sprint for important QRs 5.Sprint allocation based on multiple PBs 6.using preparation team to define QRs 7.Establish component team handling QRs of individual component 8.Establish QR specialist team 9.Use innovation and planning iteration** | |
| S312 | 1. Use artifacts: epics, features, acceptance criteria and Definition of Done to document QRs in product and sprint backlogs QR **2. Use wiki-pages** 3.Use mockups, wireframes, whiteboards and flipcharts to communicate QRs | |
| S142* | Practice to address security in DevOps (1.Good documentation and logging 2.Strong collaboration and communication 3.Enforcement of separation of roles) | |
| S143 | Practices to address security in ASD (1. Organizational (information security strategy, **raise security awarenes**s at managerial and developers level) 2.Team practices (**security training**, regular communication, adequate composition and competency of team) 3.Project practices (project security planning, customer involvement and applying misuse cases, abuser stories to handle security) | |
| S154 | Use of personas by product owners to understand users and their requirement in ASD | |
| S2 | Non-Functional Requirements Analysis Approach | Method |
| S10 | **SEAP (Security-Enhanced Agile Software Development Process)** | |
| S11 | Three tactics that support rapid and agile stability: aligning feature-based development and system decomposition, creating an architectural runway, and using matrix teams | |
| S23 | **SEAP (Security-Enhanced Agile Software Development Process)** | |
| S24 | Extended planning game to address security in XP | |
| S25 | Approach combining model driven architecture and Agile framework to deal with QRs in ASD | |
| S42 | NERV methodology (Nonfunctional Requirements Elicitation, Reasoning, and Validation in Agile Processes) | |
| S43 | **Method for Elicitation, Documentation and Validation of User Requirements (MEDoV)** | |
| S44 | **Method for Elicitation, Documentation and Validation of User Requirements (MEDoV)** | |
| S46 | EAST (extended agile security testing) | |
| S48 | Non-functional Requirements Modeling for Agile Processes (NORMAP) methodology | |
| S49 | Artifacts that enable Non-functional Requirements Modeling for Agile Processes (NORMAP) methodology | |
| S52 | Non-functional Requirements Planning (NORPLAN) | |
| S57 | Approach for Integrating security into Feature-Driven Development (FDD) | |
| S59 | Innovation requirement elicitation | |
| S64 | Evolutionary project management (Evo) method with a feature for specifying quality goals for QRs | |
| S67 | Requirements engineering methodology considering QRs at early stages of ASD | |
| S70 | Architecturally Savy Personas (ASP-Lite) approach to address QRs in ASD | |
| S73 | Use case patterns catalogue driven requirements engineering approach also addressing QRs | |
| S74 | ACRUM (Attribute-driven sCrum) which is a quality attribute driven agile development | |
| S75 | A method to add security activities in ASD with a tunable parameter controlling agile characteristic of the process. | |
| S81 | Usability-pattern based requirement-analysis method | |
| S82 | Approach to combine XP and software architecture design considering QRs | |
| S85 | InterMod to address usability in ASD | |
| S86 | secure Scrum | |
| S87 | **CEP methodology (capturing, eliciting and prioritizing NFRs) in ASD** | |
| S88 | **CEP methodology (capturing, eliciting and prioritizing NFRs) in ASD** | |
| S89 | ScrumS (an agile method aimed to improve system quality, reliability and security) | |
| S91 | CRUISER agile cross-discipline user interface and software engineering lifecycle | |
| S92 | Method for managing QRs in ASD | |
| S93 | S-Scrum (Secure Scrum) | |
| S94 | SENoR (Structured Elicitation of Non-functional Requirements) | |
| S97 | Scenario-based usability technique to manage usability in XP | |
| S99 | Method extending ASD to produce acceptably secure software | |
| S108 | Approach for the embodiment of security activity in ASD | |
| S110 | Approach to handling security and performance NFRs in projects involving big data and cloud, using Scrum | |
| S111 | XWebProcess tailored ASD method to deal with QRs | |
| S116 | U-SCRUM for addressing usability in Scrum | |
| S118 | Method for integrating security requirements n ASD | |
| S120 | Agent oriented modelling (AOM) approach using goal modeling and behavioral scenarios to elicit and represent requirements including QRs in ASD | |
| S121 | Tagging approach to elicit security requirements in ASD | |
| S134 | Q-Rapids, data driven method to elicit, assess and document QRs in ASD | |
| S135 | Proposes machine learning approach to automatically analyze user stories and identify QRs in ASD | |
| S149 | Threat poker as team based method to assess security and privacy risks in ASD | |
| S153 | Integrated approach of discount usability and UCD with scrum to address usability | |
| S155 | S-Scrum that integrates system theoretic process analysis to address safety in Scrum in safety critical systems | |
| S5 | **Agile NFR traceability model** | Model |
| S6 | Hybrid model for prioritizing requirements including QRs in Scrum | |
| S28 | Model for integrating usability into ASD | |
| S29 | Agile Usability Software Model | |
| S34 | RASP (Risk-Based, Architecture-Centric Strategic Prototyping) model | |
| S54 | Secure Feature Driven Development (SFDD) Model | |
| S55 | **Agile NFR traceability model** | |
| S60 | Enhanced Scrum model with security backlog and security master | |
| S68 | software Performance Requirements Evolution Model (PREM) to specify and validate performance requirements in ASD | |
| S83* | Modified Non-homogeneous Poisson process model to model and analyze software reliability in open source software | |
| S96* | CIViT model for visualizing end to end testing, supporting QR testing in continuous integration | |
| S100 | SMM [Story cards based Requirements engineering Maturity Model] to identify QRs and improve quality in ASD | |
| S105 | US-Scrum model (hybrid model based on Feature Driven Development (FDD) and Scrum principles to accommodate quality focus in terms of correctness, security and usability | |
| S131 | Traceability process model of ASD for Tracing NFR change impact (TANC) | |
| S5 | SAgile tool to trace NFRs in ASD | Tool |
| S13 | Declarative authorization plugin tool using Domain-specific Language (DSL) to address security in ASD | |
| S14 | Change Assistant GUI tool to support authorization in ASD | |
| S48 | **NORMATIC for visualizing and modeling NFRs in ASD** | |
| S50 | **NORMATIC for visualizing and modeling NFRs in ASD** | |
| S53* | Perflab tool to assess how new code affects performance in continuous deployment | |
| S55 | SAgile tool to trace NFRs in ASD | |
| S62 | POLVO tool to address usability in ASD | |
| S79 | SQUISH Usability automation tool in ASD | |
| S114 | J3DPerfUnit tool to support specification and testing of performance requirements in ASD | |
| S147 | NFRs recommendation system for scrum based projects | |
| S4 | UCD framework to address usability in ASD | Framework |
| S26 | **AFFINE (Agile Framework For Integrating Nonfunctional requirements Enginee)** | |
| S27 | **AFFINE (Agile Framework For Integrating Nonfunctional requirements Enginee)** | |
| S47 | RE-KOMBINE framework to support light weight requirements process in ASD | |
| S51 | NORVIEW, planning and visualization framework that would be used to schedule software non-functional requirements implementations in ASD | |
| S66 | Framework to address security in ASD | |
| S102 | Secure Scrum, an extension of the software development framework Scrum. | |
| S117 | Agile security framework | |
| S136 | Secure software development framework to address security in ASD | |
| S146 | Framework to integrate UX in ASD | |
| S150 | Architectural refactoring framework to attain required levels of NFR's through formalizing Spikes and DoD's within Scrum practices | |
| S151 | Framework to help in understanding and implementing usability in ASD | |
| S152* | Framework to support the process of (re-) architecting for continuous delivery and deployment (RSD) by considering QRs | |
| S12 | Set of guidelines for integrating Agile methods and security in Agile Projects | Guidelines |
| S38* | Catalog of DevOps patterns to scale web applications using cloud services (addressing scalability, availability and performance) | |
| S81 | Guideline to transform user tasks into a set of application features on a UI page, and then to integrate requirements analysis results with probated usability factors – usability properties and patterns | |
| S127 | Guidelines for XP to better deal with the requirements in the development of secure software | |
| S132 | Guidelines proposal for documenting QRs in ASD, according to their scope and detail | |
| S140 | Guidelines to address usability in ASD | |
| S156 | QR elicitation guidelines for ASD | |
| S3 | Suggestion for more focus on security in every step of development and increasing security awareness among stakeholder in ASD | Advices |
| S13 | Suggestions for systematic threat analysis and risk assessment to get understanding of which security aspects to focus in ASD | |
| S21 | Apply quality assurance methods to address security in AD at least twice in the development lifecycle: once after first several iterations in a project, and once closer to the end, i.e., several iterations before the system is expected to be shipped. | |
| S30 | A clear usability perspective is needed from the project management as well as the organizational context for successful integration of the user perspective in Scrum. People also need to have the skill to be able to shoulder the responsibility for usability, and to work | |
| | with usability in complex settings | |
| S37 | Product owners should have more security awareness | |
| S45 | Having an UE domain expert in the development team also assures that generic Usability Requirements are taken into account during requirement gathering activities; it is necessary to explicitly involve users on-site for certain UE activities instead of different customer stakeholders; design decisions can consider Usability Requirements and technical constraints in an easy and early stage | |
| S61 | Clarify the non-functional requirements early in the project. | |
| S84 | Make performance requirements explicit early, and plan proper levels of testing | |
| S103 | Security audits by consultants, developer trainings and security experts need to combine their security knowledge with insights of a company's organizational structures and prevailing development practices | |
| S133 | Suggest to include Safety team member to address safety in ASD | |
| S137* | Provide security education, also conducting incident response exercises in DeVops for both developers and operations | |

## References

[1] Capgemini, Sogeti, M. Focus, World Quality Report 2018-19, 2018.

[2] C. Jones, O. Bonsignour, The economics of software quality, Addison-Wesley Professional, 2011.

[3] B. Ramesh, L. Cao, R. Baskerville, Agile requirements engineering practices and challenges: an empirical study, Inf. Syst. J. 20 (2010) 449–480. doi:10.1111/j.1365-2575.2007.00259.x.

[4] W.N. Behutiye, P. Rodríguez, M. Oivo, A. Tosun, Analyzing the concept of technical debt in the context of agile software development: A systematic literature review, Inf. Softw. Technol. 82 (2017) 139–158. doi:10.1016/j.infsof.2016.10.004.

[5] B. Kitchenham, S.L. Pfleeger, Software quality: the elusive target, IEEE Softw. 13 (1996) 12–21. doi:10.1109/52.476281.

[6] K. Wiegers, J. Beatty, Software Requirements, Microsoft Press. (2013) 637. doi:978-0-7356-7966-5.

[7] K. Beck, M. Beedle, A. Van Bennekum, A. Cockburn, W. Cunningham, M. Fowler, J. Grenning, J. Highsmith, A. Hunt, R. Jeffries, J. Kern, B. Marick, R.C. Martin, S. Mellor, K. Schwaber, J. Sutherland, D. Thomas, Manifesto for Agile Software Development, Agil. Alliance. 2009 (2001) 2006. http://agilemanifesto.org/.

[8] P. Rodríguez, A. Haghighatkhah, L.E. Lwakatare, S. Teppola, T. Suomalainen, J. Eskeli, T. Karvonen, P. Kuvaja, J.M. Verner, M. Oivo, Continuous deployment of software intensive products and services: A systematic mapping study, J. Syst. Softw. 123 (2017) 263–291. doi:10.1016/j.jss.2015.12.015.

[9] P. Rodríguez, J. Markkula, M. Oivo, K. Turula, Survey on agile and lean usage in finnish software industry, in: Int. Symp. Empir. Softw. Eng. Meas., 2012: pp. 139–148. doi:10.1145/2372251.2372275.

[10] W. Alsaqaf, M. Daneva, R. Wieringa, Quality Requirements in Large-Scale Distributed Agile Projects -- A Systematic Literature Review, in: P. Grünbacher, A. Perini (Eds.), Requir. Eng. Found. Softw. Qual. 23rd Int. Work. Conf. REFSQ 2017, Essen, Ger. Febr. 27 -- March 2, 2017, Proc., Springer International Publishing, Cham, 2017: pp. 219–234. doi:10.1007/978-3-319-54045-0_17.

[11] E.-M. Schön, J. Thomaschewski, M.J. Escalona, Agile Requirements Engineering: A systematic literature review, Comput. Stand. Interfaces. 49 (2017) 79–91. doi:10.1016/j.csi.2016.08.011.

[12] S. Bartsch, Practitioners' perspectives on security in agile development, in: Proc. 2011 6th Int. Conf. Availability, Reliab. Secur. ARES 2011, 2011: pp. 479–484. doi:10.1109/ARES.2011.82.

[13] L. Cao, B. Ramesh, Agile requirements engineering practices: An empirical study, IEEE Softw. 25 (2008) 60–67. doi:10.1109/MS.2008.1.

[14] S. Bellomo, R.L. Nord, I. Ozkaya, A study of enabling factors for rapid fielding combined practices to balance speed and stability, Proc. - Int. Conf. Softw. Eng. (2013) 982–991. doi:10.1109/ICSE.2013.6606648.

[15] A. Martini, J. Bosch, The Danger of Architectural Technical Debt: Contagious Debt and Vicious Circles, in: Proc. - 12th Work. IEEE/IFIP Conf. Softw. Archit. WICSA 2015, 2015: pp. 1–10. doi:10.1109/WICSA.2015.31.

[16] K. Curcio, T. Navarro, A. Malucelli, S. Reinehr, Requirements engineering: A systematic mapping study in agile software development, J. Syst. Softw. 139 (2018) 32–50. doi:10.1016/j.jss.2018.01.036.

[17] V.T. Heikkilä, D. Damian, C. Lassenius, M. Paasivaara, A Mapping Study on Requirements Engineering in Agile Software Development, 2015 41st Euromicro Conf. Softw. Eng. Adv. Appl. (2015) 199–207. doi:10.1109/SEAA.2015.70.

[18] I. Inayat, S.S. Salim, S. Marczak, M. Daneva, S. Shamshirband, A systematic literature review on agile requirements engineering practices and challenges, Comput. Human Behav. 51 (2015) 915–929. doi:10.1016/j.chb.2014.10.046.

[19] D.A. Magües, J.W. Castro, S.T. Acuña, Requirements engineering related usability techniques adopted in agile development processes, in: Proc. Int. Conf. Softw. Eng. Knowl. Eng. SEKE, 2016: pp. 537–542. doi:10.18293/SEKE2016-057.

[20] P. Heck, A. Zaidman, A systematic literature review on quality criteria for agile requirements specifications, Softw. Qual. J. (2016). doi:10.1007/s11219-016-9336-4.

[21] K. Elghariani, N. Kama, Review on Agile requirements engineering challenges, in: 2016 3rd Int. Conf. Comput. Inf. Sci. ICCOINS 2016 - Proc., 2016. doi:10.1109/ICCOINS.2016.7783267.

[22] H. Villamizar, M. Kalinowski, M. Viana, D.M. Fernández, A systematic mapping study on security in agile requirements engineering, in: Proc. - 44th Euromicro Conf. Softw. Eng. Adv. Appl. SEAA 2018, 2018: pp. 454–461. doi:10.1109/SEAA.2018.00080.

[23] M.V.. Mäntylä, B.. Adams, F.. Khomh, E.. Engström, K.. Petersen, On rapid releases and software testing: a case study and a semi-systematic literature review, Empir. Softw. Eng. 20 (2015) 1384–1425. doi:10.1007/s10664-014-9338-4.

[24] E. Laukkanen, J. Itkonen, C. Lassenius, Problems, causes and solutions when adopting continuous delivery—A systematic literature review, Inf. Softw. Technol. 82 (2017) 55–79. doi:10.1016/j.infsof.2016.10.001.

[25] T. Karvonen, W. Behutiye, M. Oivo, P. Kuvaja, Systematic literature review on the impacts of agile release engineering practices, Inf. Softw. Technol. 86 (2017) 87–100. doi:10.1016/j.infsof.2017.01.009.

[26] C.R. Klaus Pohl, Requirements Engineering Fundamentals A Study Guide for the Certified Professional for Requirements Engineering Exam, Rocky Nook. 53 (2015) 1–184. doi:10.1017/CBO9781107415324.004.

[27] L. Chung, J.D.P. Leite, On Non-Functional Requirements in Software Engineering, Concept. Model. Found. …. (2009) 363–379. doi:10.1007/978-3-642-02463-4_19.

[28] J. Doerr, D. Kerkow, T. Koenig, T. Olsson, T. Suzuki, Non-functional requirements in industry - three case studies adopting an experience-based NFR method, in: 13th IEEE Int. Conf. Requir. Eng., 2005. doi:10.1109/RE.2005.47.

[29] J. Mylopoulos, L. Chung, B. Nixon, Representing and Using Nonfunctional Requirements: A Process-Oriented Approach, IEEE Trans. Softw. Eng. 18 (1992) 483–497. doi:10.1109/32.142871.

[30] M. Shaw, Truth vs knowledge: The difference between what a component does and what we know it does, in: Proc. 8th Int. Work. Softw. Specif. Des., 1996: p. 181.

[31] R.L. Glass, Defiling Quality Intuitively, IEEE Softw. 15 (1998) 103–104. doi:10.1109/52.676973.

[32] M. Glinz, On Non-Functional Requirements, 15th IEEE Int. Requir. Eng. Conf. (RE 2007). (2007) 21–26. doi:10.1109/RE.2007.45.

[33] B.W. Boehm, J.R. Brown, H. Kaspar, Characteristics of software quality, (1978).

[34] J.A. McCall, P.K. Richards, G.F. Walters, Factors in software quality, Natl. Tech. Inf. Serv. (1977). doi:RADC-TR-77-369/CDRL A003.

[35] G.C. Roman, TAXONOMY OF CURRENT ISSUES IN REQUIREMENTS ENGINEERING., Computer (Long. Beach. Calif). (1985). doi:10.1109/MC.1985.1662861.

[36] T.P. Bowen, G.B. Wigle, J.T. Tsai, Specification of Software Quality Attributes. Volume 3. Software Quality Evaluation Guidebook, 1985.

[37] I.S.O. Iso, Iec 9126-1: Software engineering-product quality-part 1: Quality model, Geneva, Switz. Int. Organ. Stand. 21 (2001). https://webstore.iec.ch/preview/info_isoiec9126-1%7Bed1.0%7Den.pdf.

[38] International Organization For Standardization, ISO/IEC 25010 - Systems and software engineering - Systems and software Quality Requirements and Evaluation (SQuaRE) - System and software quality models, 2011. http://www.iso.org/iso/iso_catalogue/catalogue_tc/catalogue_detail.htm?csnumber=35733.

[39] J.P. Carvallo, X. Franch, C. Quer, Managing non-technical requirements in COTS components selection, in: Proc. IEEE Int. Conf. Requir. Eng., 2006. doi:10.1109/RE.2006.40.

[40] R.B. Svensson, M. Höst, B. Regnell, Managing quality requirements: A systematic review, in: Proc. - 36th EUROMICRO Conf. Softw. Eng. Adv. Appl. SEAA 2010, 2010: pp. 261–268. doi:10.1109/SEAA.2010.55.

[41] D. Ameller, C. Ayala, J. Cabot, X. Franch, How do software architects consider non-functional requirements: An exploratory study, in: 2012 20th IEEE Int. Requir. Eng. Conf. RE 2012 - Proc., 2012: pp. 41–50. doi:10.1109/RE.2012.6345838.

[42] D. Ameller, C. Ayala, J. Cabot, X. Franch, Non-functional requirements in architectural decision making, IEEE Softw. 30 (2013) 61–67. doi:10.1109/MS.2012.176.

[43] R.R. Maiti, F.J. Mitropoulos, Capturing, eliciting, predicting and prioritizing (CEPP) non-functional requirements metadata during the early stages of agile software development, in: Conf. Proc. - IEEE SOUTHEASTCON, 2015. doi:10.1109/SECON.2015.7133007.

[44] T. Little, Context-adaptive agility: Managing complexity and uncertainty, IEEE Softw. 22 (2005) 28–35. doi:10.1109/MS.2005.60.

[45] S. Jeon, M. Han, E. Lee, K. Lee, Quality attribute driven agile development, in: Proc. - 2011 9th Int. Conf. Softw. Eng. Res. Manag. Appl. SERA 2011, 2011. doi:10.1109/SERA.2011.24.

[46] VersionOne.com, 11th Annual State of Agile Report, VersionOne Agil. Annu. Rep. (2017). doi:10.1093/jicru/ndl025.

[47] B. Fitzgerald, K.-J. Stol, Continuous software engineering and beyond: trends and challenges, in: Proc. 1st Int. Work. Rapid Contin. Softw. Eng. - RCoSE 2014, ACM Press, New York, New York, USA, 2014: pp. 1–9. doi:10.1145/2593812.2593813.

[48] T.S. Da Silva, A. Martin, F. Maurer, M. Silveira, User-centered design and agile methods: A systematic review, in: Proc. - 2011 Agil. Conf. Agil. 2011, 2011: pp. 77–86. doi:10.1109/AGILE.2011.24.

[49] J. Medeiros, M. Goulao, A. Vasconcelos, C. Silva, Towards a model about quality of software requirements specification in agile projects, in: Proc. - 2016 10th Int. Conf. Qual. Inf. Commun. Technol. QUATIC 2016, 2017: pp. 236–241. doi:10.1109/QUATIC.2016.058.

[50] J.D.R.V. Medeiros, D.C.P. Alves, A. Vasconcelos, C. Silva, E. Wanderley, Requirements engineering in agile projects: A systematic mapping based in evidences of industry, CIBSE 2015 - XVIII Ibero-American Conf. Softw. Eng. (2015) 460–473.

[51] S. Kitchenham, B. and Charters, Guidelines for performing systematic literature reviews in software engineering, in: Tech. Report, Ver. 2.3 EBSE Tech. Report. EBSE, 2007.

[52] K. Petersen, R. Feldt, S. Mujtaba, M. Mattsson, Systematic Mapping Studies in Software Engineering, in: Proc. 12th Int. Conf. Eval. Assess. Softw. Eng., British Computer Society, Swinton, UK, UK, 2008: pp. 68–77. http://dl.acm.org/citation.cfm?id=2227115.2227123.

[53] C. Wohlin, Guidelines for snowballing in systematic literature studies and a replication in software engineering, in: Proc. 18th Int. Conf. Eval. Assess. Softw. Eng. - EASE '14, 2014: pp. 1–10. doi:10.1145/2601248.2601268.

[54] V. Garousi, M. Felderer, M. V. Mäntylä, Guidelines for including grey literature and conducting multivocal literature reviews in software engineering, Inf. Softw. Technol. 106 (2019) 101–121. doi:10.1016/j.infsof.2018.09.006.

[55] M. Kuhrmann, D.M. Fernández, M. Daneva, On the pragmatic design of literature studies in software engineering: an experience-based guideline, Empir. Softw. Eng. 22 (2017) 2852–2891. doi:10.1007/s10664-016-9492-y.

[56] M. Ivarsson, T. Gorschek, A method for evaluating rigor and industrial relevance of technology evaluations, Empir. Softw. Eng. 16 (2011) 365–395. doi:10.1007/s10664-010-9146-4.

[57] I.C. Benchmark, F. Russell, C. Consultation, I. Industry, C. Structures, R.G. Sectors, F.R. Insights, Industry Classification Benchmark (Equity), FTSE Russell. (2019). https://research.ftserussell.com/products/downloads/ICB_Rules_new.pdf?_ga=2.29975551.1749455608.1569041284-1889502384.1565973875.

[58] C. Theisen, M. Dunaiski, L. Williams, W. Visser, Writing good software engineering research papers: Revisited, in: Proc. - 2017 IEEE/ACM 39th Int. Conf. Softw. Eng. Companion, ICSE-C 2017, 2017: p. 402. doi:10.1109/ICSE-C.2017.51.

[59] S. Brinkkemper, Method engineering: Engineering of information systems development methods and tools, Inf. Softw. Technol. 38 (1996) 275–280. doi:10.1016/0950-5849(95)01059-9.

[60] D.S. Cruzes, T. Dyba, Recommended Steps for Thematic Synthesis in Software Engineering, in: Empir. Softw. Eng. Meas. (ESEM), 2011 Int. Symp., 2011: pp. 275–284. doi:10.1109/ESEM.2011.36.

[61] X. Zhou, Y. Jin, H. Zhang, S. Li, X. Huang, A map of threats to validity of systematic literature reviews in software engineering, in: Proc. - Asia-Pacific Softw. Eng. Conf. APSEC, 2017: pp. 153–160. doi:10.1109/APSEC.2016.031.

[62] V. Sachdeva, L. Chung, Handling non-functional requirements for big data and IOT projects in Scrum, in: Proc. 7th Int. Conf. Conflu. 2017 Cloud Comput. Data Sci. Eng., 2017: pp. 216–221. doi:10.1109/CONFLUENCE.2017.7943152.

[63] W.M. Farid, F.J. Mitropoulos, NORMATIC: A visual tool for modeling non-functional requirements in agile processes, in: Conf. Proc. - IEEE SOUTHEASTCON, 2012. doi:10.1109/SECon.2012.6196989.

[64] D.S. Cruzes, M. Felderer, T.D. Oyetoyan, M. Gander, I. Pekaric, How is security testing done in agile teams? A cross-case analysis of four software teams, in: Lect. Notes Bus. Inf. Process., 2017: pp. 201–216. doi:10.1007/978-3-319-57633-6_13.

[65] H.F. Soares, N.S.R. Alves, T.S. Mendes, M. Mendonca, R.O. Spinola, Investigating the Link between User Stories and Documentation Debt on Software Projects, in: Proc. - 12th Int. Conf. Inf. Technol. New Gener. ITNG 2015, 2015: pp. 385–390. doi:10.1109/ITNG.2015.68.

[66] V. Garousi, K. Petersen, B. Ozkan, Challenges and best practices in industry-academia collaborations in software engineering: A systematic literature review, Inf. Softw. Technol. 79 (2016) 106–127. doi:10.1016/j.infsof.2016.07.006.

[67] J. Jabeen, Y.H. Motla, M.A. Abbasi, D.E.B. Batool, R. Butt, S. Nazir, S.A. Anwer, Incorporating artificial intelligence technique into DSDM, in: Asia-Pacific World Congr. Comput. Sci. Eng. APWC CSE 2014, IEEE, 2014: pp. 1–8. doi:10.1109/APWCCSE.2014.7053838.

[68] J. Zou, L. Xu, W. Guo, M. Yan, D. Yang, X. Zhang, Which non-functional requirements do developers focus on? An empirical study on stack overflow using topic analysis, in: IEEE Int. Work. Conf. Min. Softw. Repos., 2015: pp. 446–449. doi:10.1109/MSR.2015.60.